# Significance testing in non-sparse high-dimensional linear models

Yinchu Zhu[*1] and Jelena Bradic[†2]

[1]Lundquist College of Business, University of Oregon
[2]Department of Mathematics, University of California, San Diego

## Abstract

In high-dimensional linear models, the sparsity assumption is typically made, stating that most of the parameters are equal to zero. Under the sparsity assumption, estimation and, recently, inference have been well studied. However, in practice, sparsity assumption is not checkable and more importantly is often violated; a large number of covariates might be expected to be associated with the response, indicating that possibly all, rather than just a few, parameters are non-zero. A natural example is a genome-wide gene expression profiling, where all genes are believed to affect a common disease marker. We show that existing inferential methods are sensitive to the sparsity assumption, and may, in turn, result in the severe lack of control of Type-I error. In this article, we propose a new inferential method, named CorrT, which is robust to model misspecification such as heteroscedasticity and lack of sparsity. CorrT is shown to have Type I error approaching the nominal level for *any* models and Type II error approaching zero for sparse and many dense models. In fact, CorrT is also shown to be optimal in a variety of frameworks: sparse, non-sparse and hybrid models where sparse and dense signals are mixed. Numerical experiments show a favorable performance of the CorrT test compared to the state-of-the-art methods.

## 1 Introduction

Hypothesis testing for high-dimensional models is widely used. Statisticians usually utilize hypothesis tests to study the importance of one or many variables in parametric models. Given pairs of observations $(y_i, \mathbf{w}_i)$, with $y_i \in \mathbb{R}, \mathbf{w}_i \in \mathbb{R}^p$ and $i = 1, \ldots, n$, the importance of $j$-th variable of a parametric model $E[y_i|\mathbf{w}_i] = \mathbf{w}_i^\top \boldsymbol{\beta}$ is studied by developing hypothesis tests for the related null hypothesis $H_0 : \beta_j = 0$ against a suitable alternative, for example $H_1 : \beta_j \neq 0$. In high-dimensional settings where $p \gg n$, several theoretical results are available on the design of such tests, including asymptotic optimality [70, 32, 63, 3, 45, 16]. A common thread among these works is an underlying assumption on the sparsity of the parametric model at hand. A recent method introduced by [73], extends this body of work by removing some of the restrictive assumptions pertaining to the sparsity of the parametric model.

In particular, [73] introduced the notion of restructured regression where a specific moment of interest leads to robust and stable inference in high-dimensional setting. The authors introduce feature stabilization and synthesization pertaining to the null hypothesis of interest as a way of transforming a parametric null into a suitable and testable moment condition that is equivalent to the null hypothesis. We split into the parameter under testing $\beta^*$ and the nuisance parameter $\boldsymbol{\gamma}^*$. Applied to the problem of testing for significance of variables in high-dimensional linear models, a restructured regression as proposed in [73] takes the form of
$$\mathbf{Y} = \mathbf{X}\boldsymbol{\gamma}^* + \mathbf{Z}\beta^* + \boldsymbol{\varepsilon}, \tag{1.1}$$
where $\mathbf{W} := (\mathbf{w}_1^\top, \ldots, \mathbf{w}_n^\top)^\top = [\mathbf{X}, \mathbf{Z}] \in \mathbb{R}^{n \times (p+1)}$ with $\mathbf{Z} \in \mathbb{R}^n$ and $\mathbf{X} \in \mathbb{R}^{n \times p}$ being the design matrix, $\mathbf{Y} = (y_1, \ldots, y_n)^\top \in \mathbb{R}^n$ and $\boldsymbol{\varepsilon} \in \mathbb{R}^n$ is the error term. When interested in testing
$$H_0 : \beta^* = \beta_0, \qquad \text{vs.} \qquad H_1 : \beta^* \neq \beta_0, \tag{1.2}$$
[73] consider an auxiliary model
$$\mathbf{Z} = \mathbf{X}\boldsymbol{\theta}^* + \mathbf{u}, \tag{1.3}$$
where $\boldsymbol{\theta}^* \in \mathbb{R}^p$ is a high-dimensional and unknown parameter and $\mathbf{u} \in \mathbb{R}^n$ is the model error. [73] showcase that one can transform the original parametric null (1.2) into the moment condition
$$H_0 : E\left[(\mathbf{V} - \mathbf{X}\boldsymbol{\gamma}^*)^\top (\mathbf{Z} - \mathbf{X}\boldsymbol{\theta}^*)\right] = 0, \text{ vs. } H_1 : E\left[(\mathbf{V} - \mathbf{X}\boldsymbol{\gamma}^*)^\top (\mathbf{Z} - \mathbf{X}\boldsymbol{\theta}^*)\right] \neq 0. \tag{1.4}$$
In the above, $\mathbf{V}$ is defined as a pseudo-response,
$$\mathbf{V} := \mathbf{Y} - \mathbf{Z}\beta_0.$$

---

[*]yzhu6@oregon.edu
[†]jbradic@ucsd.edu

This paper extends the early work of [73] into a flexible method for developing asymptotically exact and optimal hypothesis tests regarding the importance of variables in high-dimensional linear models as identified through a family of moment conditions (1.4). We aim to build inference methods that inherit the desirable empirical properties of the test of [73] – such are the preservation of type I error rate and no loss in efficiency – but can be used in the wide range of linear model setting characterized by (1.1). Two of the most significant improvements over [73] are related to (conditional) heteroscedasticity and double robustness.

For example, this paper encompasses the case where the error term $\varepsilon_i$ is heteroscedastic in that the conditional variance of $\varepsilon_i$ can depend on the model features; see Condition 1. More generally, we showcase that the moment conditions of the form (1.4) can be used to provide valid and asymptotically optimal tests as well as confidence intervals, regardless whether the initial model, (1.1), or the auxiliary model, (1.3), is dense and/or heteroscedastic. An example of a possibly non-sparse auxiliary model is the one whose covariate's dependence is represented through a dense graphical model. Although both this paper and [73] deal with the inference of high-dimensional linear models with non-sparse structures, it is worth pointing out that under the more general settings (e.g., heteroscedasticity and lack of knowledge of which one of $\boldsymbol{\gamma}^*$ and $\boldsymbol{\theta}^*$ is dense), valid inference is substantially more difficult.

Section 2 gives a detailed treatment of this perspective. Moment condition like (1.4) are understood as robust moment conditions, i.e., moment conditions are written as the product of two residuals, one arising from the initial model at hand (1.1) and the other arising from an auxiliary model on covariates (1.3). Because individual residuals can have a potentially high bias, such product stabilizes inference whenever either one of the residuals is estimated accurately. This idea of correcting for bias using an auxiliary model is closely related to the work on score decorrelation, Neyman score orthogonalization, and double machine learning; see e.g., [44, 45, 14]. Important examples in high-dimensional problems include [39] and [14] for average treatment effect estimation, [4] for quantile estimation, and [61, 72, 1] for models related to time-to-event, survival analysis. These methods are quite general and can deliver valid inference with the liberty of choosing from a large class of estimators; these estimators need to have fast enough rate of convergence for both the original model and the auxiliary model. However, even consistency is quite difficult for dense models. Therefore, robust moment conditions alone are not enough to achieve valid inference for non-sparse high-dimensional models.

We seek a unified, general framework for computationally efficient and problem-specific tests. These tests are optimized for the primary objective of testing for significance of capturing a key parameter of interest, especially when the consistent estimation of the nuisance parameters is not achievable. This pushes the limit of the known optimality of doubly robust moments even in low-dimensional setting. A consistent (but not necessarily optimal) estimation was once required; now an inconsistent estimator can be allowed.

This paper addresses the challenges resulting from inconsistency by designing a set of adaptive estimators and adopting a self-normalized structure. The challenge in generalizing moment-based methods is that their success hinges on whether the adaptive estimator adequately highlights the dimensionality and heterogeneity of the model of interest. While it is natural to require estimators that achieve an optimal rate of convergence in appropriate norms (e.g., $\ell_1$-norm or prediction norm) for sparse models, achieving even consistency is not a trivial problem for dense models. Hence, we design estimators that enjoy additional adaptive properties. If the target parameter is sparse, our estimator enjoys similar properties to adaptive estimators that are designed only for sparse problems; if the target parameter is dense, our estimator has the stability needed for valid inference. When we use such adaptive estimators for both $\boldsymbol{\theta}^*$ and $\boldsymbol{\gamma}^*$, asymptotic normality holds even if only one of them is sparse. However, the correct "asymptotic variance" would depend on which one is sparse. Here, the self-normalization plays a critical role by automatically providing the correct corresponding normalization. As a result, the proposed inference method is valid without knowledge of the source of sparsity. In light of this, it is natural to see why the aforementioned existing strategies that can be coupled with generic estimators do not guarantee inference validity for dense models.

Moment conditions of the form (1.4) typically arise in scientific applications where rigorous statistical inference is required. The bulk of this paper is devoted to a theoretical analysis of moment based tests, and to establishing asymptotic normality of the resulting estimates even in the presence of severe model misspecification (e.g., non-Gaussianity, heteroscedasticity, lack of sparsity). We also develop the theory for efficiency as well as power property of the introduced tests. Our analysis is motivated by classical results for orthogonal moments, in particular [52], paired with machinery from [73] to address the adaptivity of the tests to model misspecification. The resulting framework presents a flexible method for robust statistical estimation as well as inference with formal asymptotic guarantees while allowing model misspecification.

## 1.1 Related Work

The idea of building tests on the basis of moment conditions (and orthogonal moments) has a long history, including Neyman's $C(\alpha)$ tests and Rao's score tests. They were introduced to address unobserved heterogeneity in parametric statistics models. In medicine, popular applications of these techniques include testing for rare variants [36, 42], estimation of treatment effects and causal parameters [28], in economics, for estimation and inference in missing data models as well as studying the effects of program or policy interventions [29]. At its essence the method necessitates that the moment condition used to identify the parameter of interest, needs to be insensitive towards small changes in the estimated nuisance function.

A challenge facing this approach is that if the covariate space has more than two or three dimensions performance can suffer if some of the parameters are estimated at a rate slower than $n^{-1/4}$. Unfortunately, in misspecified and high-dimensional models such rate cannot be guaranteed. For well specified models, i.e., Gaussian models, [73] showcased that this rate can be ignored. However, our paper proposes a generalization of their work in that the developed test is both robust and efficient even when the model is not correctly specified.

The original condition for Neyman orthogonality was proposed by [44], building on insights from the semi-parametric estimation framework; its optimality in linear models was established in [43]. It was then utilized to develop a class of methods called doubly robust procedures, designed to mitigate selection bias, nonrandom treatment assignment in observational studies and noncompliance in randomized experiments [53, 64]. Doubly robust methods can be viewed as a refinement of a weighted estimating-equations approach to regression with incomplete data proposed by [54, 55] and [56].

The perspective we take on moment condition (1.4) is a form of weighted estimating-equations,

where the weights are defined to be adaptive and to stabilizing for model misspecification. However, we most closely build on the proposal of [73] designed for Gaussian regression models. This adaptively weighted estimating-equations perspective also underlies several statistical analyses of missing data and causal inference problems in the context of improved local efficiency (see for example [57]).

Our adaptive weighting scheme draws heavily from a long tradition in the literature of adaptive estimation; specifically, where the dimensionality of the model is high [21, 13, 24, 18, 17]. Our goal differs from the above in that we are not focused on adaptive estimation itself. Instead, we are interested in constructing a test that is as robust as possible to model misspecification; we then rely on the self-normalization to achieve statistical stability.

In this sense, our approach is related to the studies of nonparametric function estimation where a "gap" between estimation and inference has been well established, i.e., a gap between the existence of adaptive risk bounds and the nonexistence of adaptive confidence statements. We showcase here a reciprocal "gap"; a "gap" between the existence of confidence intervals and the nonexistence of risk or estimation bounds. We argue here in this article, that optimal inference procedures exist even in the models for which estimation consistency for the model parameters cannot be guaranteed.

Our asymptotic theory relates to an extensive recent literature on the high-dimensional statistical inference, most of which focuses on the cases of Gaussian models [31, 32, 34, 12, 10]; correctly specified models [70, 63, 3, 45, 16]. Our present paper complements this body of work by showing how methods developed to study models with missing data can also be used to develop tests for misspecified models through robust and yet adaptive tests.

## 1.2 Motivating example: a challenge of the lack of sparsity

To better understand whether discoveries based on current state-of-the-art methods in high-dimensional inference can be spurious in the models that lack sparsity, we consider the problem of testing (1.2) in the model (1.1) with $\beta_0 = 0$. For emphasize the issue of lack of sparsity, we consider a homoscedastic Gaussian model. Assume the simple setup of orthogonal designs and known noise level: the entries of $\mathbf{Z}$, $\mathbf{X}$ and $\boldsymbol{\varepsilon}$ are known to be independent standard normal random variables. Moreover, assume that $\log p = o(n)$. Let $\beta^* = 0$ and $\boldsymbol{\gamma}^* = ap^{-1/2}\mathbf{1}_p$, where $a \in [-10, 10]$ is a fixed constant and $\mathbf{1}_p = (1, \ldots, 1)^\top \in \mathbb{R}^p$. It is easy to see that $\boldsymbol{\gamma}^*$ is not sparse if $a \neq 0$. Let $\boldsymbol{\pi}^* = (\beta^*, \boldsymbol{\gamma}^{*\top})^\top \in \mathbb{R}^{p+1}$ and $\mathbf{W} = (\mathbf{Z}, \mathbf{X}) \in \mathbb{R}^{n \times (p+1)}$. With $\lambda = 16\sqrt{n^{-1}\log p}$ we define the Lasso estimator $\widehat{\boldsymbol{\pi}} = \arg\min_{\boldsymbol{\pi} \in \mathbb{R}^{p+1}} \frac{1}{2n}\|\mathbf{Y} - \mathbf{W}\boldsymbol{\pi}\|_2^2 + \lambda\|\boldsymbol{\pi}\|_1$. The de-biased estimator is then defined as $\widetilde{\boldsymbol{\pi}} = \widehat{\boldsymbol{\pi}} + \widehat{\boldsymbol{\Theta}}\mathbf{W}^\top(\mathbf{Y} - \mathbf{W}\widehat{\boldsymbol{\pi}})/n$, where $\widehat{\boldsymbol{\Theta}}$ is the estimated precision matrix of the design[1]. Let $\widetilde{\beta}$ be the first entry of $\widetilde{\boldsymbol{\pi}}$. Therefore, a Wald test for the hypothesis $\beta^* = 0$ with nominal size $\alpha \in (0, 1)$ can be defined as a decision to reject the hypothesis whenever $|\widetilde{\beta}| > \Phi^{-1}(1 - \alpha/2)/\sqrt{n}$, where $\Phi(\cdot)$ is the cumulative distribution function of the standard normal distribution. The Type I error of this test is characterized in the following result.

**Theorem 1.** *In the above setup, a high-dimensional Wald test satisfies*

$$\lim_{n \to \infty} P\left(|\widetilde{\beta}| > \Phi^{-1}(1 - \alpha/2)/\sqrt{n}\right) = F(\alpha, a),$$

*where* $F(\alpha, a) = 2 - 2\Phi\left[\Phi^{-1}\left(1 - \frac{\alpha}{2}\right)/\sqrt{1 + a^2}\right]$.

It is immediately obvious that when the model is sparse, i.e., $a = 0$, we have that $F(\alpha, a) = \alpha$ and thus the Type I error of the test is asymptotically equal to the nominal level. However, when the model is not sparse, i.e., $a \neq 0$, we have that $F(\alpha, a) > \alpha$ (see Figure 1), meaning that the Type I error asymptotically exceeds the nominal level. In fact, in the non-sparse case, the Type I error of the Wald test can be close to one. In Figure 1, we see that the state-of-the-art Wald test with nominal size 1% can reject, in large samples, a true null hypothesis with probability as high as 80%. In the proof of Theorem 1, we show that $\widehat{\boldsymbol{\pi}} = 0$ with high-probability; this turns out to be a minimax optimal estimator in $\ell_2$-balls; see [19]. However, even though the zero vector resulted from (entry-wise) shrinkage methods might be a good approximation of $\boldsymbol{\pi}^*$ in the $\ell_\infty$-norm, this approximation is poor for the inference problems. Hence, this illustrates the drawback of the Wald principle in non-sparse models in general, as even optimal estimators might not be good enough for this principle to perform well.

---

[1] Since the true precision matrix is $\mathbb{I}_{p+1}$ the $(p+1) \times (p+1)$ identity matrix, we set $\widehat{\boldsymbol{\Theta}} = \mathbb{I}_{p+1}$ for simplicity; our result still holds if we use the nodewise Lasso estimator in Equation (8) of **(author?)** [63]. Moreover, observe that the method of [33] is equivalent to **(author?)** [63] in this particular setting. By Theorem 2.2 therein, $\sqrt{n}(\widetilde{\beta} - \beta^*) = \xi + o_P(1)$, where $\xi$ conditional on $\mathbf{W}$ has a normal distribution with mean zero and variance $\mathbf{Z}^\top\mathbf{Z}/n$.

Figure 1: Plot of the asymptotic Type I error of the high dimensional Wald test in a linear model with independent design. The horizontal axis denotes $a$ where $\boldsymbol{\gamma}^* = a p^{-1/2} \mathbf{1}_p$ and the vertical axis denotes the rejection probability $F(\alpha, a)$ under the null hypothesis.

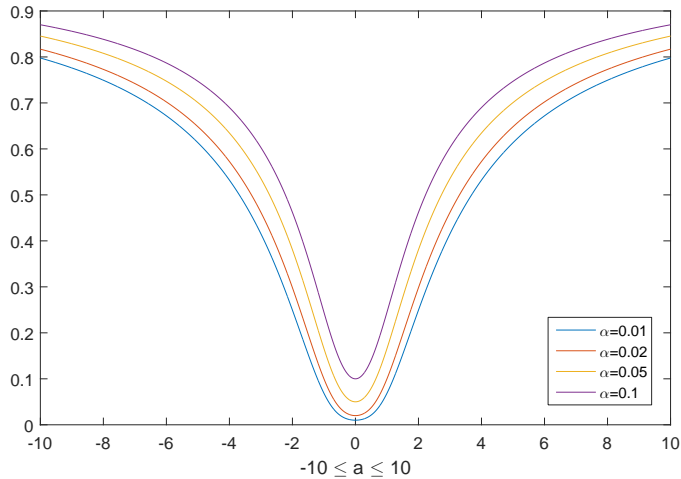

Observe that, in practice, rejecting the hypothesis that certain regression coefficient is zero is typically interpreted as evidence supporting new scientific discovery, e.g., treatment effect. However, due to the non-robustness to the lack of sparsity as demonstrated above, researchers could, with high probability, obtain "discoveries" that do not exist. Therefore, new inferential tools are called for that do not reject true hypotheses too often regardless of whether the sparsity condition holds.

### 1.3 Notation and organization

Throughout the article, we use bold upper-case letters for matrices and lower-case letters for vectors. Moreover, $^\top$ denotes the matrix transpose and $\mathbb{I}_p$ denotes the $p \times p$ identity matrix. For a vector $\mathbf{v} \in \mathbb{R}^k$, $v_j$ denotes its $j$th entry and its $\ell_q$-norm is defined as follows: $\|\mathbf{v}\|_q = (\sum_{i=1}^k |v_i|^q)^{1/q}$ for $q \in (0, \infty)$, $\|\mathbf{v}\|_\infty = \max_{1 \leq i \leq k} |v_i|$ and $\|\mathbf{v}\|_0 = \sum_{i=1}^k \mathbb{I}\{v_i = 0\}$, where $\mathbb{I}$ denotes the indicator function. For any matrix $A$, $\sigma_{\min}(A)$ and $\sigma_{\max}(A)$ denote the minimal and maximal singular values of $A$. For two sequences $a_n, b_n > 0$, we use $a_n \asymp b_n$ to denote that there exist positive constants $C_1, C_2 > 0$ such that $\forall n, a_n \leq C_1 b_n$ and $b_n \leq C_2 a_n$. We also introduce two definitions that will be used frequently. The sub-Gaussian norm of a random variable $X$ is defined as $\|X\|_{\psi_2} = \sup_{q \geq 1} q^{-1/2} (E|X|^q)^{1/q}$, whereas the sub-Gaussian norm of a random vector $\mathbf{Y} \in \mathbb{R}^k$ is $\|\mathbf{Y}\|_{\psi_2} = \sup_{\|\mathbf{v}\|_2 = 1} \|\mathbf{v}^\top \mathbf{Y}\|_{\psi_2}$. A random variable or a vector is said to be sub-Gaussian if its sub-Gaussian norm is finite. Moreover, a random variable $X$ is said to have an exponential-type tail with parameter $(b, \gamma)$ if $\forall x > 0, P(|X| > x) \leq \exp[1 - (x/b)^\gamma]$. This is a generalization of the sub-Gaussian property; see [67].

The rest of the paper is organized as follows. Section 2 discusses the main idea behind the proposed CorrT procedure. It introduces the moment construction technique and a construction of a self-normalizing test statistic related to that moment of interest. Subsection 2.1 designs tuning-adaptive estimators that are particularly useful for estimation in potentially non-sparse models. Theoretical results are presented in Section 3, including robustness to lack of sparsity, sparsity-adaptive property and relation to the oracle. Section 4 shows numerical examples and contrasts the outcomes with two state-of-the-art methods. The detailed proofs of the main results and that of a number of auxiliary results are collected in the Appendix.

## 2 Correlation Test: CorrT

In standard doubly robust methods as proposed by [44, 54], the test for a particular parameter of interest is constructed by solving the estimating equations with respect to that parameter of interest; the solution would provide a direct and consistent way of constructing confidence intervals or hypothesis tests (1.4). However, in high-dimensional setting, we abandon this approach and construct a test statistic directly mimicking the estimating equations. There is significant advantage of doing so when the model parameter or feature correlation can be extremely dense.

### 2.1 Self-adaptive and sparsity-adaptive estimation

We begin by providing estimates for the unknown parameters $\boldsymbol{\gamma}^*$ and $\boldsymbol{\theta}^*$, which will be used to construct the CorrT test.

One approach to estimating these parameters is to apply one of the many methods designed for sparse problems; e.g., the Lasso estimator, Dantzig selector, self-tuning methods by [5], [59] and [24], and many more. Although these methods often work well in sparse models, they are extremely sensitive to the model misspecification and the level of sparsity. Here, we provide adaptive and stable estimators that can be used in conjunction with (1.4). Our new algorithms introduce problem-specific constraints for sparsity-inducing regularization. This construction delivers an adaptivity in the following sense. When the estimation target fails to be sparse, the size of the estimated residuals is stable and does not diverge too quickly with the sample size due to the constraints; when the estimation target is sparse, the estimator automatically achieves consistency through the $\ell_1$-regularization.

Our method is based on the solution path of the following linear program. For $a > 0$, let

$$
\begin{aligned}
\widetilde{\boldsymbol{\gamma}}(a) \quad &:= \quad \underset{\boldsymbol{\gamma} \in \mathbb{R}^p}{\arg \min} \|\boldsymbol{\gamma}\|_1 \\
s.t. \quad & \|n^{-1}\mathbf{X}^\top(\mathbf{V} - \mathbf{X}\boldsymbol{\gamma})\|_\infty \leq \eta_0 a \\
& \|\mathbf{V} - \mathbf{X}\boldsymbol{\gamma}\|_\infty \leq \|\mathbf{V}\|_2 / \log^2 n \\
& n^{-1}\mathbf{V}^\top(\mathbf{V} - \mathbf{X}\boldsymbol{\gamma}) \geq \rho_n n^{-1}\|\mathbf{V}\|_2^2.
\end{aligned}
\tag{2.1}
$$

where the constants $\eta_0 = 1.1n^{-1/2}\Phi^{-1}(1 - p^{-1}n^{-1})$ and $\rho_n = 0.01/\sqrt{\log n}$. The first inequality in (2.1), much like the Dantzig selector, provides a necessary gradient condition guaranteeing that the gradient of the loss is close to zero. Observe that constant $a$ is different from $(E\varepsilon_i^2)^{1/2}$ as the model is heteroscedastic in nature. Optimal choice of the constant $a$ is driven by (2.2). The constant $\eta_0$ is scale-free and is derived from moderate deviation results of self-normalized sums. The second and third inequalities introduce stability in estimation, whenever the model parameter is grossly non-sparse.

Estimation of the "variance" is extremely important for any testing problem. Since we consider a heteroscedastic model, the variance of the error terms might not be informative as $E(x_{i,j}^2 \varepsilon_i^2) \neq E(x_{i,j}^2)E(\varepsilon_i^2)$. For this reason, we invoke moderate deviation results for self-normalized sums and would like $a$ to target the quantity $a_{0,\gamma}$, where

$$
a_{0,\gamma}^2 = \max_{1 \leq j \leq p} n^{-1} \sum_{i=1}^n x_{i,j}^2 \varepsilon_i^2.
$$

This quantity mimics the celebrated [69]'s heteroscedasticity-robust standard error. Hence, we treat the unknown quantity $a_{0,\gamma}$ as a parameter, which is chosen to be the largest element in the set $\mathcal{S}_\gamma$

defined as

$$\mathcal{S}_\gamma = \left\{ a \geq 0 : \quad \frac{3}{2}a \geq \sqrt{\max_{1\leq j\leq p} n^{-1} \sum_{i=1}^n x_{i,j}^2 (v_i - x_i^\top \widetilde{\gamma}(a))^2} \geq \frac{1}{2}a \right\}. \quad (2.2)$$

In other words, $\widehat{a}_\gamma = \arg\max\{a : a \in \mathcal{S}_\gamma\}$. The requirements in the set $\mathcal{S}_\gamma$ can be viewed as a data-dependent way of detecting sparsity and the tuning parameter $a$.

Our final estimator is then defined as the following combination

$$\widehat{\gamma} = \widetilde{\gamma}(\widehat{a}_\gamma)\mathbf{1}\{\mathcal{S}_\gamma \neq \emptyset\} + \widetilde{\gamma}(2\|\mathbf{X}\|_\infty \|\mathbf{V}\|_2/\sqrt{n})\mathbf{1}\{\mathcal{S}_\gamma = \emptyset\}. \quad (2.3)$$

When $\gamma^*$ is sparse, we show that $\mathcal{S}_\gamma \neq \emptyset$ with probability approaching one and that the $\ell_1$-norm in the objective function, the constraint on $\|n^{-1}\mathbf{X}^\top(\mathbf{V} - \mathbf{X}\gamma)\|_\infty$ and the definition of $\mathcal{S}_\gamma$ induce the self-adapting property by adjusting $\widehat{a}_\gamma$ to be close to $\sqrt{\max_{1\leq j\leq p} E[x_{i,j}^2 \varepsilon_i^2]}$. Thus, for sparse $\gamma^*$, $\widetilde{\gamma}(\widehat{a}_\gamma)$ is consistent and behaves like other sparsity-based estimations, such as Lasso or Dantzig selector with ideal choice of tuning parameters.

When $\gamma^*$ is not sparse, the constraints in the definition of $\widetilde{\gamma}(\cdot)$ guarantee that $\|\mathbf{X}^\top(\mathbf{V}-\mathbf{X}\widehat{\gamma})\|_\infty/\|\mathbf{V}-\mathbf{X}\widehat{\gamma}\|_2$ is not growing too fast; see Lemma 3 in the Appendix. Hence, the resulting estimator $\widehat{\gamma}$ is not only stable but also is able to automatically "detect" sparseness and exploit it to achieve estimation accuracy.

The estimation of $\theta^*$ is done in a similar manner. We define $a \mapsto \widetilde{\theta}(a)$ by

$$\begin{aligned}
\widetilde{\theta}(a) &= \arg\min_{\theta \in \mathbb{R}^p} \|\theta\|_1 \\
s.t. &\quad \|n^{-1}\mathbf{X}^\top(\mathbf{Z} - \mathbf{X}\theta)\|_\infty \leq \eta_0 a \\
&\quad \|\mathbf{Z} - \mathbf{X}\theta\|_\infty \leq \|\mathbf{Z}\|_2/\log^2 n \\
&\quad n^{-1}\mathbf{Z}^\top(\mathbf{Z} - \mathbf{X}\theta) \geq \rho_n n^{-1}\|\mathbf{Z}\|_2^2.
\end{aligned} \quad (2.4)$$

where $\eta_0$ and $\rho_n$ are as defined before. Let $\widehat{a}_\theta = \arg\max\{a : a \in \mathcal{S}_\theta\}$, where

$$\mathcal{S}_\theta = \left\{ a \geq 0 : \quad \frac{3}{2}a \geq \sqrt{\max_{1\leq j\leq p} n^{-1} \sum_{i=1}^n x_{i,j}^2 (z_i - x_i^\top \widetilde{\theta}(a))^2} \geq \frac{1}{2}a \right\}.$$

The final estimator is then defined as

$$\widehat{\theta} = \widetilde{\theta}(\widehat{a}_\theta)\mathbf{1}\{\mathcal{S}_\theta \neq \emptyset\} + \widetilde{\theta}(2\|\mathbf{X}\|_\infty \|\mathbf{Z}\|_2/\sqrt{n})\mathbf{1}\{\mathcal{S}_\theta = \emptyset\}. \quad (2.5)$$

### 2.2 CorrT Test

We now introduce the test statistic. The main difference between the proposed test related to other high-dimensional tests is that its asymptotic distribution is not affected by misspecification (e.g., in terms of heteroscedasticity and sparsity) of models (1.1) and (1.3). Motivated by the success of the test proposed by [73], our approach mimics the plug-in nature of their test while automatically tailoring normalization of the test to the appropriate level.

Let $\widehat{\gamma}$ and $\widehat{\theta}$ be defined in (2.3) and (2.5), respectively. We propose to consider the following correlation test (CorrT) statistic

$$T_n(\beta_0) = \frac{\widehat{\varepsilon}^\top \widehat{\mathbf{u}}}{\sqrt{\sum_{i=1}^n \widehat{\varepsilon}_i^2 \widehat{u}_i^2}}, \quad (2.6)$$

where $\widehat{\varepsilon} = \mathbf{V} - \mathbf{X}\widehat{\gamma}$, $\widehat{\mathbf{u}} = \mathbf{Z} - \mathbf{X}\widehat{\theta}$. The notation $T_n(\beta_0)$ indicates that the test statistic is constructed with the knowledge of the null hypothesis $H_0 : \beta = \beta_0$. Whenever possible we suppress its dependence on $\beta_0$ and use $T_n$ instead.

Self-normalizing nature of the test statistics also has an advantage that it allows us to derive an analytical critical value for the test; we show that $T_n(\beta_0)$, under the null hypothesis, converges in distribution to $N(0,1)$. Hence, a test with nominal size $\alpha \in (0,1)$ rejects the hypothesis (1.2) if and only if $|T_n(\beta_0)| > \Phi^{-1}(1-\alpha/2)$.

We briefly discuss the mechanism of our test and outline the reason for its robustness to the lack of sparsity in $\boldsymbol{\gamma}^*$ or $\boldsymbol{\theta}^*$; in fact the method is blind to the choice of the sparse parameter and provides valid inference when either of the two models (1.1) and (1.3) is sparse. The product structure in (2.6) helps to decouple the bias introduced by $\widehat{\boldsymbol{\gamma}}$ or $\widehat{\boldsymbol{\theta}}$, allowing us to establish uncorrelatedness between the estimated residuals $\widehat{\boldsymbol{\varepsilon}}$ and $\mathbf{u}$ under the null hypothesis $H_0$. We can show, without assuming sparsity of $\boldsymbol{\gamma}^*$, that

$$n^{-1/2}\widehat{\boldsymbol{\varepsilon}}^\top \widehat{\mathbf{u}} = n^{-1/2}\widehat{\boldsymbol{\varepsilon}}^\top \mathbf{u} + O_P(\sqrt{\log p}\|\widehat{\boldsymbol{\theta}} - \boldsymbol{\theta}^*\|_1).$$

Under the null hypothesis, the first term on the right hand side has zero expectation and the second term vanishes fast enough due to sparse $\boldsymbol{\theta}^*$. On the other hand, without assuming sparsity of $\boldsymbol{\theta}^*$, we can show that

$$n^{-1/2}\widehat{\boldsymbol{\varepsilon}}^\top \widehat{\mathbf{u}} = n^{-1/2}\boldsymbol{\varepsilon}^\top \widehat{\mathbf{u}} + O_P(\sqrt{\log p}\|\widehat{\boldsymbol{\gamma}} - \boldsymbol{\gamma}^*\|_1).$$

Under the null hypothesis, the first term on the right hand side has zero expectation and the second term vanishes fast enough due to sparse $\boldsymbol{\gamma}^*$.

The construction of $\widehat{\boldsymbol{\gamma}}$ and $\widehat{\boldsymbol{\theta}}$ as in (2.3) and (2.5) is fundamentally critical to establishing limiting distribution of the leading term in the decomposition above. We note that commonly used estimators in sparse regression of $\mathbf{Y}$ against $\mathbf{X}$ and $\mathbf{Z}$ do not deliver such de-correlation and, for non-sparse $\boldsymbol{\gamma}^*$ or $\boldsymbol{\theta}^*$, do not possess tractable properties.

### 2.3 Computational Aspects: parametric simplex method

The proposed method allows for efficient implementation. We point out that both optimization problems (2.1) and (2.4) can be written in the form of a parametric linear program, where the "additional unknown parameter", $a$, is present only as an upper bound of the linear constraints. It is well-known that, if a problem can be cast as a parametric linear program, then the parametric simplex method can be used to obtain the optimal solution. Computational burden of obtaining the solution paths $a \mapsto \widetilde{\boldsymbol{\gamma}}(a)$ is the same of the burden of computing the solution for one value of $a$; see [66] and [47]. We now explicitly formalize the optimization problem (2.1) as parametric linear programs; the formulation of (2.4) is analogous. Let $a > 0$, and denote with $\widetilde{\boldsymbol{\gamma}}(a) = \widehat{\mathbf{b}}^+(a) - \widehat{\mathbf{b}}^-(a)$, where $\widehat{\mathbf{b}}(a) = (\widehat{\mathbf{b}}^+(a)^\top, \widehat{\mathbf{b}}^-(a)^\top)^\top$ is defined as the solution to the following parametric right hand side linear program

$$\widehat{\mathbf{b}}(a) = \underset{\mathbf{b} \in \mathbb{R}^{2p}}{\arg\max}\, \mathbf{M}_1^\top \mathbf{b} \qquad \text{subject to} \qquad \mathbf{M}_2 \mathbf{b} \leq \mathbf{M}_3 + \mathbf{M}_4 a, \tag{2.7}$$

where the matrices $\mathbf{M}_1 \in \mathbb{R}^{2p \times 1}, \mathbf{M}_2 \in \mathbb{R}^{(2p+2n+1) \times 2p}, \mathbf{M}_3 \in \mathbb{R}^{2p+2n+1}, \mathbf{M}_4 \in \mathbb{R}^{2p+2n+1}$ are taken to be $\mathbf{M}_1 = -\mathbf{1}_{2p \times 1}$,

$$\mathbf{M}_2 = \begin{pmatrix} n^{-1}\mathbf{X}^\top \mathbf{X} & -n^{-1}\mathbf{X}^\top \mathbf{X} \\ -n^{-1}\mathbf{X}^\top \mathbf{X} & n^{-1}\mathbf{X}^\top \mathbf{X} \\ \mathbf{X} & -\mathbf{X} \\ -\mathbf{X} & \mathbf{X} \\ n^{-1}\mathbf{V}^\top \mathbf{X} & -n^{-1}\mathbf{V}^\top \mathbf{X} \end{pmatrix}, \mathbf{M}_3 = \begin{pmatrix} n^{-1}\mathbf{X}^\top \mathbf{V} \\ -n^{-1}\mathbf{X}^\top \mathbf{V} \\ \mathbf{V} + \mathbf{1}_{n \times 1}\|\mathbf{V}\|_2/\log^2 n \\ -\mathbf{V} + \mathbf{1}_{n \times 1}\|\mathbf{V}\|_2/\log^2 n \\ (1-\rho_n)n^{-1}\mathbf{V}^\top \mathbf{V} \end{pmatrix}, \mathbf{M}_4 = \begin{pmatrix} \eta_0 \mathbf{1}_{p \times 1} \\ \eta_0 \mathbf{1}_{p \times 1} \\ \mathbf{0}_{n \times 1} \\ \mathbf{0}_{n \times 1} \\ 0 \end{pmatrix}$$

It is known that the solution of a parametric simplex algorithm is a piecewise linear and convex function of $a$.

## 3 Theoretical Properties

In this section we present theoretical properties of the proposed method and consider an asymptotic regime in which $p$ and the sparsity level of $\boldsymbol{\gamma}^*$ or $\boldsymbol{\theta}^*$ are allowed to grow to infinity faster than $n$.

Suppose that we have $n$ independent and identically distributed (i.i.d) observations, indexed $i = 1, \ldots, n$. For each observation, we have access to the response variable $y_i$ along with a set of auxiliary covariates $\mathbf{w}_i = (z_i, \mathbf{x}_i^\top)^\top \in \mathbb{R}^{p+1}$. Here, $\mathbf{x}_i$, $z_i$ and $y_i$ denote the $i$-th row of $\mathbf{X}$, the $i$-th entry of $\mathbf{Z}$ and of $\mathbf{Y}$, respectively. The $j$-th entry of $\mathbf{x}_i$ will be denoted by $x_{i,j}$. Similar notations will be used for $u_i$, $v_i$ and $\varepsilon_i$, which are the $i$-th entry of $\mathbf{u}$, $\mathbf{V}$ and $\boldsymbol{\varepsilon}$, respectively.

In order to establish the theoretical claims, we impose the following regularity condition.

**Condition 1.** *There are constants $\kappa_1, \ldots, \kappa_5 > 0$ such that the following hold. (i) Let $\boldsymbol{\Sigma}_W = E[\mathbf{w}_1 \mathbf{w}_1^\top] \in \mathbb{R}^{p \times p}$ satisfy $\kappa_1 \leq \sigma_{\min}(\boldsymbol{\Sigma}_W) \leq \sigma_{\max}(\boldsymbol{\Sigma}_W) \leq \kappa_2$ and $\|\mathbf{w}_i\|_{\psi_2} \leq \kappa_2$. (ii) In addition, $0 < 1/\kappa_3 < E(\varepsilon_i^2) < \kappa_3$, $E(z_i \varepsilon_i) = 0$ and $E(\mathbf{x}_i \varepsilon_i) = 0$. (iii) $E(u_i \mid \mathbf{x}_i, \varepsilon_i) = 0$, $P(\kappa_4 < \sigma_{u,i}^2 < \kappa_5) = 1$ and $\|u_i\|_{\psi_2} \leq \kappa_5$, where $\sigma_{u,i}^2 = E(u_i^2 \mid \mathbf{x}_i, \varepsilon_i)$.*

A few comments are immediate. For generality, we set up Condition 1 in an abstract way. We notice that in the context of our main problems of interest requiring Condition 1 is not particularly stringent. Unlike [73] or [31] – among references that allow dense model parameters – we require only sub-Gaussian designs.

In addition, we consider a non-sparse model that allows for conditional heteroscedasticity. If the errors $\varepsilon_i$ are being considered as independent of the features, non-sparse models were studied by [73]. However, it has been observed (e.g. [40, 3]) that heteroscedasticity could have severe consequences for high-dimensional models. In particular, the Lasso or Dantzig selector become very unstable in this context. Heteroscedasticity in this paper refers to conditional heteroscedasticity, which means that the conditional variance of $\varepsilon_i$ depends on the features. A natural example is that of $\varepsilon_i = \sigma(\mathbf{x}_i) \xi_i$, where $\sigma(\cdot)$ is a measurable function and $\xi_i$ is independent of $\mathbf{x}_i$ with $E(\xi_i) = 0$ and $E(\xi_i^2) = 1$. Even for low-dimensional models, heteroscedasticity has been known to cause several complications (for example, Gauss-Markov theorem no longer applies) with a large literature on constructing valid inference in the presence of heteroscedasticity, see e.g., [48, 26, 69].

Conditional heteroscedasticity is also allowed in the auxiliary model, therefore allowing for feature heteroscedasticity; for example, variance of one feature may directly depend on many other features in the model. This includes a wide spectrum of dependence structures among the features.

Our next assumption controls the size of the two models.

**Condition 2.** *For some constant $\kappa_6 > 0$, $\|\boldsymbol{\gamma}^*\|_2 \leq \kappa_6$ and $s_\theta = o\left(n^{1/2}/(\log(p \vee n))^{5/2}\right)$, where $s_\theta = \|\boldsymbol{\theta}^*\|_0$.*

The rate for $s_\theta$ is slightly stronger than the conditions in [3] and [45] who impose $o(\sqrt{n}/\log p)$ and in [63] who impose $o(n/\log p)$. However, we do not impose that the original model is sparse as long as a row of the precision matrix $\boldsymbol{\theta}^*$ is sparse enough. This can be interpreted as the cost of allowing for non-sparse $\boldsymbol{\gamma}^*$ as well as general heteroscedatic model errors. Moreover, the validity of CorrT is also guaranteed for dense $\boldsymbol{\theta}^*$ and sparse $\boldsymbol{\gamma}^*$; see Theorem 3. Therefore, our test can detect automatically which one of the two ($\boldsymbol{\gamma}^*$ or $\boldsymbol{\theta}^*$) is sparse and utilize it for effective inference; hence, the practitioner does not need knowledge of which of the two is sparse when applying our method.

Additionally, in studying the estimation problem for signals with many nonzero entries, it is common to consider parameters in bounded convex balls, e.g. [20], [51] and [31]. Notice that the bounded $\ell_2$-norm itself does not impose direct constraints on the sparsity, see e.g., [31].

### 3.1 Size properties: validity regardless of sparsity

Given these assumptions, we are now ready to provide an asymptotic characterization of the proposed CorrT test.

**Theorem 2.** *Let Conditions 1 and 2 hold. Then under the null hypothesis (1.2),*

$$\forall \alpha \in (0, 1), \quad \lim_{n, p \to \infty} P\left(|T_n(\beta_0)| > \Phi^{-1}(1 - \alpha/2)\right) = \alpha, \tag{3.1}$$

*where $T_n(\beta_0)$ is defined in Equation (2.6).*

Theorem 2 formally establishes that the new CorrT test is asymptotically exact in testing $\beta^* = \beta_0$ while allowing $p \gg n$. In particular, CorrT is robust to dense $\boldsymbol{\gamma}^*$ or $\boldsymbol{\theta}^*$, in the sense that even under dense $\boldsymbol{\gamma}^*$ or $\boldsymbol{\theta}^*$, the proposed procedure does not generate false positive results. Compared to existing methods, we have moved away from the unverifiable model sparsity assumption and hence can handle more realistic problems.

**Remark 1.** *Our result is theoretically intriguing as it circumvents limitations of the "inference based on estimation" principle. This principle relies on accurate estimation, which is challenging, if not impossible, for non-sparse and high-dimensional models. To see the difficulty, consider the full model parameter $\boldsymbol{\pi}^* := (\beta^*, \boldsymbol{\gamma}^{*\top})^\top \in \mathbb{R}^{p+1}$. The minimax rate of estimation in terms of $\ell_2$-loss for parameters in a $(p+1)$-dimensional $\ell_q$-ball with $q \in [0,1]$ and of radius $r_n$ is $r_n(n^{-1}\log p)^{1-q/2}$ (see [51]). Theorem 2 says that CorrT is valid even when $\boldsymbol{\pi}^*$ cannot be consistently estimated. For example, suppose that $\log p \asymp n^c$ for a constant $c > 0$ and $\boldsymbol{\pi}_j^* = 1/\sqrt{p}$ for $j = 1, \ldots, p+1$. Then, as $n \to \infty$, $\|\boldsymbol{\pi}^*\|_q (n^{-1}\log p)^{1-q/2} \to \infty$ for any $q \in [0,1]$, suggesting potential failure in estimation.*

Instead of solely relying on estimation of $\boldsymbol{\pi}^*$, we impose the null hypothesis in our inference procedure and fully exploit its implication. With sophistication in constructing the test, the inaccuracy in $\boldsymbol{\pi}^*$ does not impair the validity (Type I error control) of our approach.

With an almost analogous argument as in the proof of Theorem 2, we can show the following result.

**Theorem 3.** *Let Conditions 3 and 4 in Appendix A hold. Then under the null hypothesis (1.2),*

$$\forall \alpha \in (0,1), \quad \lim_{n,p \to \infty} P\left(|T_n(\beta_0)| > \Phi^{-1}(1 - \alpha/2)\right) = \alpha, \tag{3.2}$$

*where $T_n(\beta_0)$ is defined in Equation (2.6).*

By Theorem 3, the asymptotic validity of CorrT still holds if we replace $s_\theta$ with $s_\gamma := \|\boldsymbol{\gamma}^*\|_0$ in the statement of Condition 2. This means that as long as one of $\boldsymbol{\gamma}^*$ and $\boldsymbol{\theta}^*$ is sparse, CorrT delivers valid inference. In this sense, CorrT automatically detects and adapts to sparse structures from different sources in the data, either in the model parameter or the covariance matrix of features; prior knowledge of the exact source is not needed.

We note that confidence sets for $\beta^*$ can be constructed by inverting the CorrT test. Let $1 - \alpha$ be the nominal coverage level. We define

$$\mathcal{C}_{1-\alpha} := \left\{\beta : |T_n(\beta)| \leq \Phi^{-1}(1 - \alpha/2)\right\}. \tag{3.3}$$

Theorem 2 guarantees the validity of this confidence set.

**Corollary 4.** *Under Conditions 1 and 2, the confidence set $\mathcal{C}_{1-\alpha}$, (3.3), has the exact coverage asymptotically,*

$$\lim_{n,p \to \infty} P\left(\beta^* \in \mathcal{C}_{1-\alpha}\right) = 1 - \alpha.$$

*Moreover, the same conclusion holds if we replace Conditions 1 and 2 with Conditions 3 and 4.*

Corollary 4 implies that the confidence set (3.3) provides exact asymptotic coverage even when the nuisance parameter $\boldsymbol{\gamma}^*$ or $\boldsymbol{\theta}^*$ is non-sparse in that $s_\gamma/p \to 1$ or $s_\theta/p \to 1$ with $p \gg n$. To the best of our knowledge, this result is unique in the existing literature. As we step further into the non-sparse regime (as we take more and more entries to be none-zero), existing results could provide confidence sets with coverage probability far below the nominal level. In the example discussed in Section 1.2 with $s_\gamma = p$, Figure 1 indicates that the 99% confidence interval based on debiasing method can have coverage probability around 20%.

We would also like to point out that Theorem 2 holds uniformly over a large parameter space and over a range of null hypotheses. To formally state the result, we define the nuisance parameter $\xi^* = (\gamma^*, \theta^*, F)$, where $F$ is the distribution of $(\mathbf{x}_i, \varepsilon_i, u_i)$. Notice that the distribution of the observed data $\{(y_i, x_i, z_i)\}_{i=1}^n$ is determined by $(\beta^*, \xi^*)$. The probability distribution under $(\beta^*, \xi^*)$ is denoted by $P_{(\beta^*, \xi^*)}$. The space for the nuisance parameter we consider is

$$\Xi(s_\theta) = \{\xi = (\gamma, \theta, F) : \text{ Condition 1 holds with } \kappa_1, \ldots, \kappa_5 > 0, \ \|\gamma\|_2 \leq \kappa_6 \text{ and } \|\theta\|_0 \leq s_\theta\},$$

where $\kappa_1, \ldots, \kappa_6 > 0$ are constants. We have the following result.

**Corollary 5.** *If $s_\theta = o\left(\sqrt{n/(\log(p \vee n))^5}\right)$, then $\forall \alpha \in (0,1)$,*

$$\limsup_{n,p \to \infty} \sup_{(\beta,\xi) \in \mathbb{R} \times \Xi(s_\theta)} \left| P_{(\beta,\xi)} \left( |T_n(\beta)| > \Phi^{-1}(1 - \alpha/2) \right) - \alpha \right| = 0.$$

Corollary 5 says that Theorem 2 holds uniformly in a large class of distributions. An analogous result can be stated for Theorem 3.

## 3.2 Power properties

In this subsection, we investigate the power properties of CorrT. Let us observe that it is still unclear what statistical efficiency means for inference in non-sparse high-dimensional models. Various minimax results on both estimation [51] and inference [30] suggest that no procedures are guaranteed to accurately identify non-sparse parameters. In light of these results, it appears that, for dense high-dimensional models, it is quite difficult, if possible at all, to obtain tests that are powerful uniformly over a large class of distributions, e.g., a bounded $\ell_2$ ball for $\gamma^*$. However, our test does have desirable power properties for certain classes of models of practical interest. First, we show that our test is efficient for sparse models. Second, we show that our test has the optimal detection rate $O(n^{-1/2})$ for important dense models.

### 3.2.1 Adaptivity: efficiency under sparsity

In Section 1.2, we have demonstrated that the lack of robustness could cause serious problems for existing inference methods designed for sparse problems. A certain oracle procedure with the knowledge of $s_\gamma = \|\gamma^*\|_0$ would proceed as follows: (1) use existing sparsity-based methods for sparse models in order to achieve efficiency in inference and (2) resort to certain conservative approaches to retain validity for dense problems. However, the sparsity level $s_\gamma$ is rarely known in practice. Hence, an important question is whether or not it is possible to design a procedure that can automatically adapt to the unknown sparsity level $s_\gamma$ and match the above oracle procedure in the following sense. We say that a procedure for testing the hypothesis (1.2) is sparsity-adaptive if (i) this procedure does not require knowledge of $s_\gamma$, (ii) provides valid inference under any $s_\gamma$ and (iii) achieves efficiency with sparse $\gamma^*$.

In this subsection, we show that the CorrT test possesses such sparsity-adaptive property. It is clear from Section 2 that CorrT does not require knowledge of $s_\gamma$. In Section 3.1, we show that CorrT provides valid inference without any assumption on $s_\gamma$. We now show the third property, efficiency under sparse $\gamma^*$. To formally discuss our results, we consider testing $H_0 : \beta^* = \beta_0$ versus

$$H_{1,h} : \ \beta^* = \beta_0 + h/\sqrt{n}. \tag{3.4}$$

where $h \in \mathbb{R}$ is a fixed constant. $H_{1,h}$ in (3.4) is called a Pitman alternative and is typically used to assess the asymptotic efficiency of tests; see [65] and [37] for classical treatments on local power.

**Theorem 6.** *Let Conditions 1 and 2 hold together with $\log p = o(\sqrt{n})$. Let $\mathbf{\Sigma}_X = E[\mathbf{x}_i \mathbf{x}_i^\top] \in \mathbb{R}^{(p-1) \times (p-1)}$. Suppose that $\|\gamma^*\|_0 = o(\sqrt{n}/\log^2(p \vee n))$, $\min_{1 \leq j \leq p} E[x_{1,j}^2 \varepsilon_1^2] \geq \tau$ and $Eu_1^2/\sqrt{E\varepsilon_1^2 u_1^2} \to \kappa$ for some constants $\tau, \kappa > 0$. Then, under $H_{1,h}$ in (3.4),*

$$\lim_{n,p \to \infty} P_{\beta^*}\left( |T_n(\beta_0)| > \Phi^{-1}(1 - \alpha/2) \right) = \Psi(h, \kappa, \alpha),$$

*where $\Psi(h, \kappa, \alpha) = \Phi\left(-\Phi^{-1}(1 - \alpha/2) + h\kappa\right) + \Phi\left(-\Phi^{-1}(1 - \alpha/2) - h\kappa\right)$.*

Theorem 6 establishes the power properties of the proposed CorrT test.

**Remark 2.** *In the case of homoscedastic errors with $\sigma_\varepsilon^2 = E\varepsilon_1^2$ and $\sigma_u^2 = Eu_1^2$, we can compare the local power of CorrT with that of existing methods. In particular we consider the local power of [63] and note that similar analysis applies to [3] and [45]. Let $\widehat{b}_{Lasso}$ denote the debiased Lasso estimator defined in [63]. Under homoscedasticity, our condition of $Eu_1^2/\sqrt{E\varepsilon_1^2 u_1^2} \to \kappa$ translates into $\sigma_u/\sigma_\varepsilon \to \kappa$. Applying Theorem 2.3 of [63] to our setup, we obtain $\sqrt{n}(\widehat{b}_{Lasso} - \beta^*) \to^d N(0, \kappa^{-2})$. Hence, a natural test, referred to as [63] test, is to reject $H_0$ in (1.2) if and only if $|\widehat{b}_{Lasso} - \beta_0| > \kappa^{-1}\Phi^{-1}(1-\alpha/2)/\sqrt{n}$. Under $H_{1,h}$ (3.4), the asymptotic normality of $\widehat{b}_{Lasso}$ implies that $\sqrt{n}(\widehat{b}_{Lasso} - \beta_0) \to^d N(h, \kappa^{-2})$. Hence, it is not hard to see that the power of [63] test against $H_{1,h}$ (3.4) is asymptotically equal to $\Psi(k, \kappa, \alpha)$ defined in Theorem 6. Since $\widehat{b}_{Lasso}$ is shown to be a semi-parametrically efficient estimator for $\beta^*$, CorrT is asymptotically equivalent to tests based on efficient estimators.*

Moreover, results from [33] suggest that our test is also minimax optimal whenever the model is homoscedastic. By Theorem 2.3 therein (adapted to the Gaussian setting), a minimax optimal $\alpha$ level test for testing $\beta^* = \beta_0$ against $H_{1,h}$ in (3.4) has power at most

$$\Phi\left(a_n h \sigma_u \sigma_\varepsilon^{-1} - \Phi^{-1}(1-\alpha/2)\right) \qquad (3.5)$$
$$+ \Phi\left(-a_n h \sigma_u \sigma_\varepsilon^{-1} - \Phi^{-1}(1-\alpha/2)\right) + F_{n-s_\gamma+1}(n - s_\gamma + l_n).$$

In the display above $l_n = \sqrt{n} \log n$, $a_n = \sqrt{(n - s_\gamma + l_n)/n}$ and $F_k(x) = P(W_k \geq x)$, where $W_k$ denotes a random variable from $\chi^2(k)$, the chi-squared distribution with $k$ degrees of freedom. By the Bernstein's inequality applied to the sum of i.i.d $\chi^2(1)$ random variables, one can easily show that $F_{n-s_\gamma+1}(n - s_\gamma + l_n) = o(1)$. Hence, as $n \to \infty$, the limit of the bound in (3.5) is equal to $\Psi(k, \kappa_0, \alpha)$ defined in Theorem 6.

However, Theorem 6 also holds under a heteroscedastic model (1.1). Observe that the only condition regarding heteroscedasticity is related to the second moment between the design and the errors, summarized by $\kappa$. Hence, the local power only depends on $\kappa$ even though heteroscedasticity allows a very rich class of dependence between $\varepsilon_i$ and $u_i$.

Theorems 2 and 6 establish the sparsity-adaptive property discussed at the beginning of Section 3.2.1: CorrT is shown to automatically detect sparse structures in $\boldsymbol{\gamma}^*$ and utilize them to optimize power, while maintaining validity even in the absence of sparsity.

### 3.2.2 Extremely dense models

One important class of non-sparse models is what we shall refer to as the "extremely dense models". In these models, a potentially important case is that the entries of the model are all small individually but are strong collectively as the dimension of the model explodes. In our testing problem (1.2), this translates into an extremely dense nuisance parameter $\boldsymbol{\gamma}^*$ with $\|\boldsymbol{\gamma}^*\|_0 = p \gg n$. Consistent estimation of such extremely dense nuisance parametrs might not be possible. However, we are able to discuss the power of the proposed test in the presence of a particular dense nuisance parameter and obtain a strong result indicating that the proposed test controls Type II errors extremely well. To the best of our knowledge, the next result is unique.

**Theorem 7.** *Let Conditions 1 and 2 hold together with $\log p = o(\sqrt{n})$. Let $\boldsymbol{\Sigma}_X = E[\mathbf{x}_i \mathbf{x}_i^\top] \in \mathbb{R}^{(p-1)\times(p-1)}$. Suppose that $\min_{1 \leq j \leq p} \text{Var}\left[x_{1,j}(\mathbf{x}_1^\top \boldsymbol{\gamma}^* + \varepsilon_1)\right] \geq \tau$, $\|\boldsymbol{\Sigma}_X \boldsymbol{\gamma}^*\|_\infty \leq \sqrt{\tau(2n)^{-1} \log(p \vee n)}/25$ and $Eu_1^2/\sqrt{E(\mathbf{x}_1^\top \boldsymbol{\gamma}^* + \varepsilon_1)^2 u_1^2} \to \kappa$ for some constants $\tau, \kappa > 0$. Then, under $H_{1,h}$ in (3.4),*

$$\lim_{n,p \to \infty} P_{\beta^*}\left(|T_n(\beta_0)| > \Phi^{-1}(1-\alpha)\right) = \Psi(h, \kappa, \alpha),$$

*where $\Psi(h, \kappa, \alpha)$ is defined in Theorem 6.*

We can draw a few conclusions from Theorem 7. For $n, p \to \infty$ with $\sqrt{\log p}/n = o(1)$, the Type II error of the proposed CorrT test, against alternatives with deviations larger than $O(n^{-1/2})$, converges to zero.

For the case of independent columns of the design matrix and the case of Toeplitz designs, the condition $\|\boldsymbol{\Sigma}_X \boldsymbol{\gamma}^*\|_\infty = O(\sqrt{(\log p)/n})$ is satisfied if $\|\boldsymbol{\gamma}^*\|_\infty = O(\sqrt{(\log p)/n})$ as long as $\boldsymbol{\gamma}^*$ lies in a bounded $\ell_2$-ball. More generally, $\|\boldsymbol{\gamma}^*\|_\infty = O(\sqrt{(\log p)/n})$ is a sufficient condition under any covariance matrix $\boldsymbol{\Sigma}_X$ satisfying $\max_{1 \leq j \leq p} \|\boldsymbol{\Sigma}_{X,j}\|_1 = O(1)$ (a condition sometimes imposed for consistent estimation of covariance matrices).

Theorem 7 offers new insight into the inference of dense models. When we test the full parameter $(\boldsymbol{\beta}^*, \boldsymbol{\gamma}^*)$ against an alternative, the minimax test might not have power against "fixed alternative" (deviation bounded away from zero) if the parameter is not sparse; see [30]. Theorem 7 says that the problem of testing single entries of a potentially dense model is an entirely different problem; we can indeed have power against alternatives of the order $O(n^{-1/2})$, just like the case of fixed $p$. Hence, this rate is optimal. This would also imply that any confidence intervals computed by inverting the test $T_n$ are optimal regardless of the size of sparsity of $\boldsymbol{\gamma}^*$.

### 3.2.3 Hybrid high-dimensional model

To showcase wide applicability of our method and to provide the bridge between sparse and dense models, we can consider what is perhaps a more practical scenario for the nuisance parameter $\boldsymbol{\gamma}^*$. Under the so-called "sparse + dense" model, $\boldsymbol{\gamma}^*$ is the sum of a sparse vector and an extremely dense vector. The estimation of high-dimensional models with this hybrid sparsity are discussed by [15], who derive bounds for prediction errors. However, their results only concern with estimation, and thus the statistical inference of these models even in moderately high dimensions is still an open problem. The following result offers the first solution to this issue.

**Theorem 8.** *Let Conditions 1 and 2 hold together with $\log p = o(\sqrt{n})$. Let $\boldsymbol{\Sigma}_X = E[\mathbf{x}_i \mathbf{x}_i^\top] \in \mathbb{R}^{(p-1) \times (p-1)}$. Suppose that $\boldsymbol{\gamma}^* = \boldsymbol{\pi}^* + \boldsymbol{\mu}^*$ for $\boldsymbol{\pi}^*$ and $\boldsymbol{\mu}^*$ such that (1) $\|\boldsymbol{\pi}^*\|_0 = o(\sqrt{n}/\log^{5/2}(p \vee n))$, (2) $(\boldsymbol{\pi}^* + \boldsymbol{\mu}^*)^\top \boldsymbol{\Sigma}_X \boldsymbol{\mu}^* + E(\varepsilon_1^2) \geq \tau_1$, (3) $\min_{1 \leq j \leq p} Var[x_{1,j}(\mathbf{x}_1^\top \boldsymbol{\mu}^* + \varepsilon_1)] \geq \tau_2$, (4) $\|\boldsymbol{\Sigma}_X \boldsymbol{\mu}^*\|_\infty \leq \sqrt{\tau_2 (2n)^{-1} \log(p \vee n)}/25$ and (5) $Eu_1^2/\sqrt{E(\mathbf{x}_1^\top \boldsymbol{\mu}^* + \varepsilon_1)^2 u_1^2} \to \kappa$, where $\tau_1, \tau_2, \kappa > 0$ are constants. Then, under $H_{1,h}$ in (3.4)*,

$$\lim_{n,p \to \infty} P_{\boldsymbol{\beta}^*}\left(|T_n(\beta_0)| > \Phi^{-1}(1 - \alpha)\right) = \Psi(h, \kappa, \alpha),$$

*where $\Psi(h, \kappa, \alpha)$ is defined in Theorem 6.*

Theorem 8 says that when high-dimensional models have parameters with the hybrid structure, testing single entries can have power in detecting deviations larger than $O(n^{-1/2})$, the optimal rate even for low-dimensional models.

Condition (4) of Theorem 8 is a natural restriction on the interaction between the sparse part and the dense part of the model. When $\boldsymbol{\Sigma}_X = \mathbb{I}_p$, condition (4) is automatically satisfied if $\boldsymbol{\pi}^*$ and $\boldsymbol{\mu}^*$ have disjoint supports. Since $(\boldsymbol{\pi}^* + \boldsymbol{\mu}^*)^\top \boldsymbol{\Sigma}_X \boldsymbol{\mu}^* + E(\varepsilon_1^2) = \|\boldsymbol{\Sigma}_X^{1/2}(\boldsymbol{\pi}^* + \boldsymbol{\mu}^*/2)\|_2^2 + E(\varepsilon_1^2) - \|\boldsymbol{\Sigma}_X^{1/2} \boldsymbol{\mu}^*\|_2^2/4$, a sufficient condition for condition (4) is $\|\boldsymbol{\Sigma}_X^{1/2} \boldsymbol{\mu}^*\|_2^2 \leq 2E(\varepsilon_1^2)$.

## 4 Numerical Examples

In this section we evaluate the finite-sample performance of the proposed method via Monte Carlo simulations. We also illustrate CorrT using a dataset on breast cancer trials.

### 4.1 Simulation Examples

We simulate the linear model $\mathbf{Y} = \mathbf{W}\boldsymbol{\pi}^* + \boldsymbol{\varepsilon}$ as follows. The design matrix $\mathbf{W} \in \mathbb{R}^{n \times p}$ consists of i.i.d rows from one of the following distributions:

**LTD** Light-tailed design: $N(0, \boldsymbol{\Sigma}_{(\rho)})$ with the $(i,j)$ entry of $\boldsymbol{\Sigma}_{(\rho)}$ being $\rho^{|i-j|}$.

**HTD** Heavy-tailed design: each row has the same distribution as $\boldsymbol{\Sigma}_{(\rho)}^{1/2}\mathbf{U}$, where $\mathbf{U} \in \mathbb{R}^p$ contains i.i.d random variables with Student's t-distribution with 3 degrees of freedom (the third moment does not exist.)

We consider both uncorrelated designs ($\rho = 0$) and the correlated designs ($\rho = -1/2$). The error term $\boldsymbol{\varepsilon} \in \mathbb{R}^n$ is drawn as a vector of i.i.d random variables from either $N(0,1)$ (light-tailed error, or LTE) or Student's t-distribution with 3 degrees of freedom (heavy-tailed error, or HTE). Let $s = \|\boldsymbol{\pi}^*\|_0$ denote the size of the model sparsity and we vary simulations settings from extremely sparse $s = 3$ to extremely large $s = p$. We set the model parameters as follows: $\boldsymbol{\pi}_j^* = 2/\sqrt{n}$ for $2 \leq j \leq 4$, $\boldsymbol{\pi}_j^* = 0$ for $j > \max\{s, 4\}$ and other entries of $\boldsymbol{\pi}^*$ are i.i.d random variables uniformly distributed on $(0, 4/\sqrt{n})$. We test the hypothesis $H_0 : \boldsymbol{\pi}_3^* = 2/\sqrt{n}$ and the rejection probability represents the Type I error.

We compare our method with the debiasing method of [63] and the score algorithm of [45]. For both approaches, we choose the scaled Lasso to estimate $\boldsymbol{\pi}^*$ and treat the true precision matrix and variance of the model as known. In a certain sense these procedures are quasi-oracle as they use part of the information on the (typically unknown) true probability distribution of the data. We do this primarily to reduce the arbitrariness of choosing tuning parameters (and hence the quality of inference) in implementing these methods. Moreover, these methods should not be expected to behave better under their original setting with unknown $\boldsymbol{\Sigma}_{(\rho)}$ and noise level than under our "ideal setting". All the rejection probabilities are computed using 100 random samples. All the tests we consider have a nominal size of 5%.

We collect the result in Table 1, where we clearly observe instability of the competing debiasing and score methods. Their Type-I error diverges steadily away from 5% with growing sparsity $s$; in extreme cases, Type I error rate could reach 50%, close to a random guess. In contrast, CorrT remains stable even when the model sparsity level is equal to $p$. This is still true as we change the correlation among the features and the thickness of tails in the distribution of the designs and errors.

We also investigate the power properties of testing $H_0 : \boldsymbol{\pi}_3^* = 2/\sqrt{n}$. The data is generated by the same model as in Table 1, except that the true value of $\boldsymbol{\pi}_3$ is now $\boldsymbol{\pi}_3^* = 2/\sqrt{n} + h/\sqrt{n}$. Table 2 presents full power curves with various values of $h$, which measures the magnitude of deviations from the null hypothesis. Hence the far left presents Type I error ($h = 0$) whereas other points on the curves correspond to Type II errors ($h \neq 0$). In all the power curves, we use the design matrices have a Toeplitz covariance matrix $\boldsymbol{\Sigma}_{(\rho)}$ with $\rho = -1/2$. The two plots in the first row of Table 1 correspond to extremely sparse models with $s = 3$: LTD + LTE structure on the left and LTD + HTE on the right. In both figures, we observe that CorrT compares equally to the existing methods. The second row corresponds to the case of $s = n$; we clearly observe that the CorrT outperforms both debiasing and score tests by providing firm Type I error and also reaching full power quickly. Lastly, we present power curves for $s = p$; for these extremely dense models, debiasing and score methods can have Type I error close to 50% whereas CorrT still provides valid inference. CorrT still achieves full power against alternatives that are $1/\sqrt{n}$ away from the null.

## 4.2 An application to transNOAH breast cancer trial data

We now use real data to illustrate the proposed method. The sample was selected from the transNOAH breast cancer trial (GEO series GSE50948), available for download at "http://www.ncbi.nlm.nih.gov/geo/query/acc.cgi?ac Genome-wide gene expression profiling was performed using micro RNA from biopsies of 114 pretreated patients with HER2+ breast cancer. The dataset contains gene expression values of about 20000 genes located on different chromosomes.

Clinically, breast cancer is classified into hormone receptor (HR)-positive, HER2+ and triple-negative breast cancer. The "HER2-positive" subtype of breast cancer over-expresses the human epidermal growth factor receptor 2 (HER2). BRCA1 is a human tumor suppressor gene that is normally expressed in the cells of breast and other tissues, where they help repair damaged DNA. Research suggests that the BRCA1 proteins regulate the activity of other genes including tumor suppressors and regulators of the cell division cycle. However, the association between BRCA1 and other genes is not completely understood. Moreover, it is believed that BRCA1 may regulate pathways that remove the damages in DNA introduced by the certain drugs. Thus understanding associations between BRCA1 and other genes provides a potentially important tool for tailoring chemotherapy in cancer treatment. HER2+ breast cancer is biologically heterogeneous and developing successful treatment is highly important. We apply the method developed in Section 2 to this dataset with the goal of testing particular associations between BRCA1 and a few other genes conditional on all remaining genes present in the study. Simply applying Lasso procedure results in an estimate of all zeros.

Table 1: Average Type I error of the CorrT test, Debias (De-biasing test) and Score test over 100 repetitions with $n = 200$ and $p = 500$. Rows represent different sparsity levels, whereas columns represent different design and error setting.

|  | LTD + LTE, $\rho = 0$ | | | LTD + LTE, $\rho = -\frac{1}{2}$ | | | HTD + LTE, $\rho = 0$ | | |
|---|---|---|---|---|---|---|---|---|---|
|  | CorrT | Debias | Score | CorrT | Debias | Score | CorrT | Debias | Score |
| $s = 1$ | 0.05 | 0.05 | 0.05 | 0.06 | 0.03 | 0.06 | 0.04 | 0.14 | 0.09 |
| $s = 3$ | 0.02 | 0.03 | 0.03 | 0.16 | 0.05 | 0.18 | 0.07 | 0.17 | 0.09 |
| $s = 5$ | 0.02 | 0.02 | 0.02 | 0.08 | 0.06 | 0.07 | 0.03 | 0.05 | 0.03 |
| $s = 10$ | 0.07 | 0.09 | 0.10 | 0.09 | 0.08 | 0.11 | 0.07 | 0.08 | 0.10 |
| $s = 20$ | 0.05 | 0.11 | 0.10 | 0.05 | 0.06 | 0.12 | 0.08 | 0.13 | 0.11 |
| $s = 50$ | 0.04 | 0.13 | 0.11 | 0.08 | 0.16 | 0.23 | 0.10 | 0.25 | 0.18 |
| $s = 100$ | 0.05 | 0.27 | 0.24 | 0.02 | 0.19 | 0.16 | 0.07 | 0.28 | 0.20 |
| $s = n$ | 0.03 | 0.36 | 0.37 | 0.16 | 0.34 | 0.35 | 0.05 | 0.33 | 0.37 |
| $s = p$ | 0.05 | 0.57 | 0.56 | 0.07 | 0.42 | 0.47 | 0.04 | 0.54 | 0.55 |
|  | LTD + HTE, $\rho = 0$ | | | LTD + HTE, $\rho = -\frac{1}{2}$ | | | HTD + HTE, $\rho = 0$ | | |
|  | CorrT | Debias | Score | CorrT | Debias | Score | CorrT | Debias | Score |
| $s = 1$ | 0.03 | 0.01 | 0.02 | 0.06 | 0.05 | 0.04 | 0.04 | 0.06 | 0.05 |
| $s = 3$ | 0.06 | 0.06 | 0.05 | 0.11 | 0.03 | 0.11 | 0.09 | 0.15 | 0.08 |
| $s = 5$ | 0.04 | 0.06 | 0.06 | 0.08 | 0.07 | 0.08 | 0.04 | 0.05 | 0.05 |
| $s = 10$ | 0.04 | 0.02 | 0.02 | 0.09 | 0.06 | 0.12 | 0.02 | 0.12 | 0.02 |
| $s = 20$ | 0.09 | 0.08 | 0.08 | 0.04 | 0.05 | 0.09 | 0.07 | 0.16 | 0.11 |
| $s = 50$ | 0.03 | 0.18 | 0.21 | 0.05 | 0.11 | 0.08 | 0.02 | 0.17 | 0.16 |
| $s = 100$ | 0.03 | 0.21 | 0.21 | 0.05 | 0.22 | 0.19 | 0.07 | 0.29 | 0.21 |
| $s = n$ | 0.07 | 0.38 | 0.36 | 0.05 | 0.20 | 0.24 | 0.05 | 0.35 | 0.33 |
| $s = p$ | 0.05 | 0.59 | 0.61 | 0.04 | 0.44 | 0.41 | 0.09 | 0.52 | 0.54 |

Table 2: Power properties of the CorrT test across different sparsity settings in high dimensional linear models. We consider extremely sparse, $s = n$ and $s = p$ cases presented through top to bottom row. We also consider design setting with light tailed and heavy tailed distributions, left to right columns. We compare CorrT (red) with de-biased (green) and score test (blue).

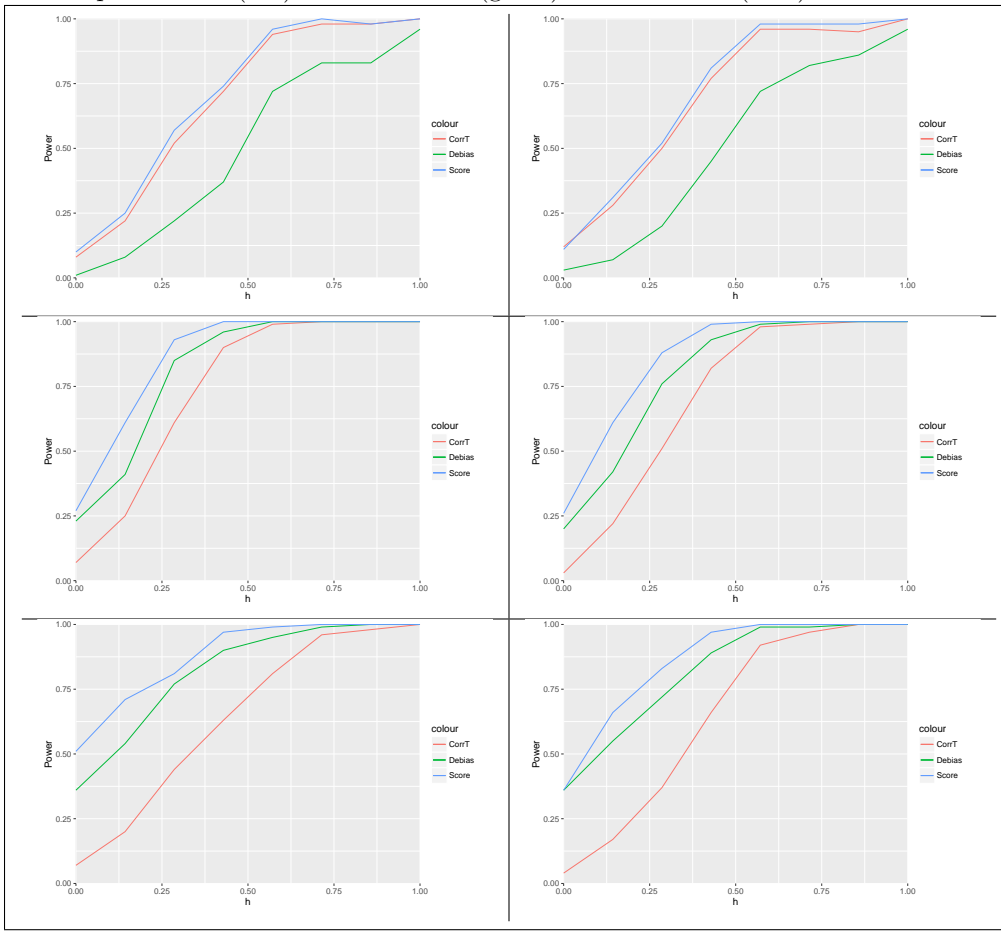

Results are reported in Table 3. Therein we report the test statistic of the proposed test, CorrT, as well as the debiasing test and the score test. We observe that there are a number of genes where the three tests report largely different values – namely CorrT reports a non-significant finding whereas the debiasing and the score test report significant findings. These genes include B3GALNT1, C3orf62, TNFAIP1 and LTB gene that have been previously linked to the lung cancer [8, 68, 71, 50], as well as CCPG1, LRRIQ3 and LOC100507537 linked to prostate, colorectal and bladder cancer, respectively [60, 2, 62]. Such findings would indicate that this dataset likely does not follow a sparse model and that previous methods might have reported false positives. Observe that even mutations related to diseases other than cancer, such as ataxia and ELOVL4 gene, are being reported as significant by the debiasing and the score tests but are found insignificant by CorrT.

Chromatin regulators have been known to act as a guide for cancer treatment choice [35]. Here we identify retinoblastoma binding protein RBBP4, a chromatin modeling factor, as being highly associated with BRCA1 in all HER2+ patients. This indicates that early detection and chemoterapy treatment should target RBBP4.

Results of Table 3 demonstrate that a partial or complete loss of NARS2 is a significant event that can be associated with the HER2+ breast cancer. Significance of this gene was confirmed in [27] where a significant correlation was found between the gene CCND1 (a known trigger of the ER positive breast cancer) and NARS2. Our analysis enriches these findings and shows statistically significant association between NARS2 and BRCA1 for HER2+ cancer cells.

Lastly, Table 3 showcases significance of a known breast tumor suppressor gene IGF2R [46]. Mannose 6-phosphate/IGF2 receptor (M6P/IGF2R) is mutated in breast cancer, resulting in loss of IGF2 degradation [22] and leading to its usage as a sensitivity marker for radiation, chemotherapy, and endocrine therapy. Germ-line mutations in BRCA1 predispose individuals to breast and ovarian cancers. A novel endogenous association of BRCA1 with Nmi (N-Myc-interacting protein) in breast cancer cells was detected in [38]. Nmi was found to interact specifically with BRCA1, both in-vitro and in-vivo, by binding to two major domains in BRCA1. Table 3 showcases significance of the association between Nmi and BRCA1 gene.

In the above analysis, our method confirms existing knowledge and also delivers new discoveries in medicine. This provides evidence that our methods can be useful for scientific research involving high-dimensional datasets.

## 5 Conclusion

In summary, our procedure achieves the optimal detection rate while adapting to any level of sparsity or lack thereof. Therefore CorrT is extremely useful for applied scientific studies where a priori knowledge of sparsity level is unknown. Since there are currently no procedures for testing the sparsity assumption CorrT is exceptionally practical. It possesses the regime-encompassing property in that it behaves optimally in all three regimes: sparse, dense and hybrid and does so automatically without user interference while allowing heteroscedastic errors, therefore, providing a comprehensive tool for significance testing.

Table 3: Value of the Test Statistic of transNOAH breast cancer trial data set concerning 114 HER+ cancer patients with 20000 genes mapped.

| Gene | Biological association | Test Statistic | | |
|---|---|---|---|---|
| | | CorrT | Debias | Score |
| IGF2R | breast cancer tumor suppressor | -4.692 | -4.285 | -4.445 |
| Nmi | endogenously associated with BRCA1 | -4.239 | -2.956 | -2.669 |
| RBBP4 | breast cancer | -4.186 | -3.314 | -2.806 |
| NARS2 | breast cancer | -4.163 | -5.000 | -4.983 |
| B3GALNT1 | lung cancer | 1.151 | 2.082 | 2.065 |
| C3orf62 | lung cancer | -1.274 | -2.143 | -2.139 |
| LTB | lung cancer | -0.131 | -2.107 | -2.143 |
| TNFAIP1 | lung cancer | 1.231 | 2.181 | 2.118 |
| CCPG1 | prostate cancer | -1.597 | -2.154 | -2.251 |
| LRRIQ3 | colorectal cancer | -1.025 | -2.480 | -2.240 |
| LOC100507537 | bladder cancer | -0.137 | -1.966 | -1.135 |
| ELOVL4 | ataxia | -1.354 | -2.152 | -2.136 |

# Appendix

## A  Assumptions for Theorem 3

**Condition 3.** *There are constants $\kappa'_1, ..., \kappa'_5 > 0$ such that the following hold. (i) $\kappa'_1 \leq \sigma_{\min}(\Sigma_W) \leq \sigma_{\max}(\Sigma_W) \leq \kappa'_2$ and $\|\mathbf{w}_i\|_{\psi_2} \leq \kappa'_2$. (ii) In addition, $0 < 1/\kappa'_3 < E(u_i^2) < \kappa'_3$, $E(z_i\varepsilon_i) = 0$ and $E(\mathbf{x}_i\varepsilon_i) = 0$. (iii) $E(\varepsilon_i \mid \mathbf{x}_i, u_i) = 0$, $P(\kappa'_4 < \sigma^2_{\varepsilon,i} < \kappa'_5) = 1$ and $\|\varepsilon_i\|_{\psi_2} \leq \kappa'_5$, where $\sigma^2_{\varepsilon,i} = E(\varepsilon_i^2 \mid \mathbf{x}_i, u_i)$.*

**Condition 4.** *For some constant $\kappa'_6 > 0$, $\|\boldsymbol{\theta}^*\|_2 \leq \kappa_6$ and $s_\gamma = o\left(n^{1/2}/(\log(p \vee n))^{5/2}\right)$, where $s_\gamma = \|\boldsymbol{\gamma}^*\|_0$.*

## B  Proofs of theoretical results

### B.1  Proof of Theorem 1

*Proof of Theorem 1.* The proof proceeds in two steps. We first show that $P(\widehat{\boldsymbol{\theta}} = 0) \to 1$ and then prove the desired result.

**Step 1: Prove that $P(\widehat{\boldsymbol{\theta}} = 0) \to 1$.**

We show that, with probability approaching one, the vector of all zeros, satisfies the KKT condition of the Lasso optimization problem (see Lemma 2.1 of **(author?)** [11]), i.e.,

$$P\left(\|n^{-1}\mathbf{W}^\top \mathbf{Y}\|_\infty \leq 16\sqrt{n^{-1}\log p}\right) \to 1. \tag{B.1}$$

Let $\xi_{i,j} = ap^{-1/2}\sum_{l=1, l\neq j}^{p} x_{i,l} + \varepsilon_i$, where $x_{i,l}$ denotes the $l$-th entry of $\mathbf{x}_i$. Notice that
(1) $\{x_{i,j}\}_{i=1}^n$ and $\{\xi_{i,j}\}_{i=1}^n$ are independent by construction of the model (design have i.i.d. rows)
,
(2) $y_i = ap^{-1/2}x_{i,j} + \xi_{i,j}$ by the definition of the model and $\xi_{i,j}$ and (3) $E\xi_{i,j}^2 = a^2(1 - p^{-1}) + 1$ for $\forall 1 \leq i \leq n$ and $\forall 1 \leq j \leq p$.

Notice that $x_{i,j}^2$ is a chi-squared random variable with one degree of freedom. Since $x_{i,j}^2 - 1$ is a chi-squared random variable right it has bounded sub-exponential norm, Proposition 5.16 of **(author?)** [67] and the union bound imply that for some constants $c_1, c_2 > 0$ and for any $z > 0$

$$P\left(\max_{1 \leq j \leq p} \left|n^{-1/2} \sum_{i=1}^n (x_{i,j}^2 - 1)\right| > z\sqrt{\log p}\right)$$

$$\leq \sum_{j=1}^p P\left(\left|n^{-1/2} \sum_{i=1}^n (x_{i,j}^2 - 1)\right| > z\sqrt{\log p}\right)$$

$$\leq 2p \exp\left[-c_1 \min\left\{\frac{z^2 \log p}{c_2^2}, \frac{z\sqrt{\log p}}{c_1 n^{-1/2}}\right\}\right].$$

It follows that there exists a constant $c_3 > 0$ such that

$$P(\mathcal{A}_n) \to 1 \quad \text{with} \quad \mathcal{A}_n = \left\{\max_{1 \leq j \leq p} \left|n^{-1} \sum_{i=1}^n x_{i,j}^2 - 1\right| > c_3 \sqrt{n^{-1} \log p}\right\}. \tag{B.2}$$

Since $\{x_{i,j}\}_{i=1}^n$ is independent of $\{\xi_{i,j}\}_{i=1}^n$, (see comment (1) above) we have that $n^{-1/2} \sum_{i=1}^n x_{i,j} \xi_{i,j}$ conditional on $\{x_{i,j}\}_{i=1}^n$ is Gaussian with mean zero and variance $n^{-1} \sum_{i=1}^n x_{i,j}^2 \sigma_\xi^2$, where $\sigma_\xi^2 = a^2(1 - p^{-1}) + 1$.

Let $w > 0$ satisfy $2 \log w = (1.45^2 \log p) / \left(1 + c_3\sqrt{n^{-1} \log p}\right)$. Then, on the event $\mathcal{A}_n$,

$$P\left(\left|n^{-1/2} \sum_{i=1}^n x_{i,j} \xi_{i,j}\right| > 1.45 \sigma_\xi \sqrt{\log p} \bigg| \{x_{i,j}\}_{i=1}^n\right)$$

$$= P\left(\frac{\left|n^{-1/2} \sum_{i=1}^n x_{i,j} \xi_{i,j}\right|}{\sigma_\xi \sqrt{n^{-1} \sum_{i=1}^n x_{i,j}^2}} > \frac{1.45\sqrt{\log p}}{\sqrt{n^{-1} \sum_{i=1}^n x_{i,j}^2}} \bigg| \{x_{i,j}\}_{i=1}^n\right)$$

$$= 1 - \Phi\left(\frac{1.45\sqrt{\log p}}{\sqrt{n^{-1} \sum_{i=1}^n x_{i,j}^2}}\right)$$

$$\stackrel{(i)}{\leq} 1 - \Phi\left(\frac{1.45\sqrt{\log p}}{\sqrt{1 + c_3\sqrt{n^{-1} \log p}}}\right)$$

$$\stackrel{(ii)}{=} 1 - \Phi\left(\sqrt{2 \log w}\right).$$

$$\stackrel{(iii)}{\leq} w^{-1}$$

$$\stackrel{(iv)}{=} \exp\left(-\frac{1.45^2 \log p}{2\left(1 + c_3\sqrt{n^{-1} \log p}\right)}\right),$$

where $(i)$ holds by the definition of $\mathcal{A}_n$, $(ii)$ holds by the definition of $w$, $(iii)$ holds by Lemma 1 and $(iv)$ holds by the definition of $w$.

By the law of iterated expectations and the union bound, we have that

$$P\left(\max_{1 \leq j \leq p} \left|n^{-1/2} \sum_{i=1}^n x_{i,j} \xi_{i,j}\right| > 1.45 \sigma_\xi \sqrt{\log p}\right) \tag{B.3}$$

$$\leq p \exp\left(\frac{-1.45^2 \log p}{2\left(1 + c_3\sqrt{n^{-1} \log p}\right)}\right) + P(\mathcal{A}_n^c) \stackrel{(i)}{=} o(1), \tag{B.4}$$

where $(i)$ holds by $p \to \infty$, $n^{-1} \log p \to 0$ and (B.2). Notice that

$$P\left(\|n^{-1/2} \mathbf{W}^\top \mathbf{Y}\|_\infty > 1.5\sqrt{(a^2+1)\log p}\right)$$

$$\stackrel{(i)}{=} P\left(\max_{1 \leq j \leq p} \left| n^{-1/2} \sum_{i=1}^n x_{i,j} \xi_{i,j} + ap^{-1/2} n^{-1/2} \sum_{i=1}^n x_{i,j}^2 \right| > 1.5\sqrt{(a^2+1)\log p}\right)$$

$$\leq P\left(\max_{1 \leq j \leq p} \left| n^{-1/2} \sum_{i=1}^n x_{i,j} \xi_{i,j} \right| > 1.45 \sigma_\xi \sqrt{\log p}\right)$$

$$+ P\left(\max_{1 \leq j \leq p} \left| ap^{-1/2} n^{-1/2} \sum_{i=1}^n x_{i,j}^2 \right| > \left(1.5\sqrt{a^2+1} - 1.45\sigma_\xi\right) \sqrt{\log p}\right)$$

$$\stackrel{(ii)}{=} o(1) + P\left(\max_{1 \leq j \leq p} \left| n^{-1} \sum_{i=1}^n x_{i,j}^2 \right| > a^{-1} \sqrt{p/n} \left(1.5\sqrt{a^2+1} - 1.45\sigma_\xi\right) \sqrt{\log p}\right)$$

$$\stackrel{(iii)}{=} o(1),$$

where $(i)$ holds by $y_i = ap^{-1/2} x_{i,j} + \xi_{i,j}$, $(ii)$ holds by (B.4) and $(iii)$ holds by (B.2) and

$$\sqrt{p/n}\left(1.5\sqrt{a^2+1} - 1.45\sigma_\xi\right) \sqrt{\log p} \asymp \sqrt{p(\log p)/n} \to \infty.$$

Since $1.5\sqrt{a^2+1} \leq 16 \; \forall a \in [-10, 10]$, we have proved (B.1). Hence, $P(\widehat{\boldsymbol{\theta}} = 0) \to 1$.

**Step 2: Establish the desired result.** Observe that $Ez_i^2 y_i^2 = Ey_i^2 = \|\boldsymbol{\gamma}^*\|_2^2 + 1 = a^2 + 1$. Then, by the central limit theorem

$$n^{-1/2} \sum_{i=1}^n z_i y_i \to^d N(0, a^2+1).$$

Since $P(\widehat{\boldsymbol{\theta}} = 0) \to 1$ (by step 1), $\sqrt{n}(\widetilde{\beta} - \beta^*) = n^{-1/2} Z^\top Y = n^{-1/2} \sum_{i=1}^n z_i y_i$ with probability approaching one. Hence,

$$\sqrt{n}(\widetilde{\beta} - \beta^*) \to^d N(0, a^2+1).$$

Since $(\widehat{\boldsymbol{\Theta}} \widehat{\boldsymbol{\Sigma}}_W \widehat{\boldsymbol{\Theta}})_{1,1} = n^{-1} \mathbf{Z}^\top \mathbf{Z} = n^{-1} \sum_{i=1}^n z_i^2$, the law of large numbers implies that $(\widehat{\boldsymbol{\Theta}} \widehat{\boldsymbol{\Sigma}}_W \widehat{\boldsymbol{\Theta}})_{1,1} = 1 + o_P(1)$. By Slutzsky's lemma, we have

$$\frac{\sqrt{n}(\widetilde{\beta} - \beta^*)}{\sqrt{(\widehat{\boldsymbol{\Theta}} \widehat{\boldsymbol{\Sigma}}_W \widehat{\boldsymbol{\Theta}})_{1,1}}} \to^d N(0, a^2+1). \tag{B.5}$$

Notice that

$$P(\beta^* \in CI_{1-\alpha}) = P\left(-\Phi^{-1}\left(1 - \frac{\alpha}{2}\right) \leq \frac{\sqrt{n}(\widetilde{\beta} - \beta^*)}{\sqrt{(\widehat{\boldsymbol{\Theta}} \widehat{\boldsymbol{\Sigma}}_W \widehat{\boldsymbol{\Theta}})_{1,1}}} \leq \Phi^{-1}\left(1 - \frac{\alpha}{2}\right)\right)$$

$$\stackrel{(i)}{=} \Phi\left(\Phi^{-1}\left(1 - \frac{\alpha}{2}\right)/\sqrt{a^2+1}\right) - \Phi\left(-\Phi^{-1}\left(1 - \frac{\alpha}{2}\right)/\sqrt{a^2+1}\right)$$

$$\stackrel{(ii)}{=} 2\Phi\left(\Phi^{-1}\left(1 - \frac{\alpha}{2}\right)/\sqrt{a^2+1}\right) - 1,$$

where $(i)$ holds by (B.5) and $(ii)$ holds by the identity $\Phi(-z) = 1 - \Phi(z)$ for $z \geq 0$. The proof is complete. $\square$

## B.2 Proof of Theorem 2

The proof of Theorem 2 is a consequence of a series of lemmas presented below.

**Lemma 1.** *For any $w \geq 5$, $\Phi^{-1}(1 - w^{-1}) \leq \sqrt{2 \log w}$. For any $w \geq 14$, $\Phi^{-1}(1 - w^{-1}) \geq \sqrt{\log w}$.*

*Proof of Lemma 1.* By Lemma 2 on page 175 of [23], we have that for any $z > 0$

$$\frac{(z^{-1} - z^{-3}) \exp(-z^2/2)}{\sqrt{2\pi}} \leq 1 - \Phi(z) \leq \frac{\exp(-z^2/2)}{\sqrt{2\pi} z}. \tag{B.6}$$

Fix an arbitrary $w \geq 5$. Since $\sqrt{2\pi} \cdot 1 \cdot \exp(1^2/2) < 5$, the continuity of $z \mapsto \sqrt{2\pi} z \exp(z^2/2)$ implies that there exists $z_0 \geq 1$ with $w = \sqrt{2\pi} z_0 \exp(z_0^2/2)$. By (B.6), we have $1 - \Phi(z_0) \leq w^{-1}$. In other words,

$$\Phi^{-1}\left(1 - w^{-1}\right) \leq z_0 \stackrel{(i)}{<} \sqrt{2 \log w},$$

where $(i)$ holds by $w = \sqrt{2\pi} z_0 \exp(z_0^2/2) > \exp(z_0^2/2)$ (since $z_0 \geq 1$). The first result follows.

To see the second result, we observe the elementary inequality $x^{-1} - x^{-3} \geq x^{-2}$ for any $x \geq (\sqrt{5} + 1)/2$. Hence, (B.6) implies that for $z \geq (\sqrt{5} + 1)/2$,

$$1 - \Phi(z) \geq \frac{(z^{-1} - z^{-3}) \exp(-z^2/2)}{\sqrt{2\pi}} \geq \frac{1}{\sqrt{2\pi} z^2 \exp(z^2/2)}.$$

In other words, for $z \geq (\sqrt{5} + 1)/2$,

$$\Phi^{-1}\left(1 - \frac{1}{\sqrt{2\pi} z^2 \exp(z^2/2)}\right) \geq z.$$

Now let $w = \exp(z^2)$. Then $z = \sqrt{\log w}$ and we have that for $w \geq 14 > \exp[(\sqrt{5}+1)^2/4]$,

$$\Phi^{-1}\left(1 - \frac{1}{\sqrt{2\pi}\sqrt{w} \log w}\right) \geq \sqrt{\log w}.$$

It is straight-forward to verify that for any $w \geq 14$, $1/(\sqrt{2\pi}\sqrt{w} \log w) \geq 1/w$ and hence

$$\Phi^{-1}\left(1 - \frac{1}{w}\right) \geq \Phi^{-1}\left(1 - \frac{1}{\sqrt{2\pi}\sqrt{w} \log w}\right).$$

The second result follows from the above two displays. The proof is complete. □

**Lemma 2.** *Let Condition 1 hold. Suppose that $H_0$ in (1.2) holds. Then, with probability tending to one, $\gamma^*$ lies in the feasible set of the optimization problem (2.1) for $a = a_{*,\gamma}$, where $a_{*,\gamma}^2 = n^{-1} \max_{1 \leq j \leq p} \sum_{i=1}^n x_{i,j}^2 \varepsilon_i^2$.*

*Proof of Lemma 2.* Under $H_0$ in (1.2), $\mathbf{Y} - \mathbf{Z}\beta_0 = \mathbf{X}\gamma^* + \varepsilon$. Then, it suffices to verify the following claims:

(a) $P\left(\|n^{-1}\mathbf{X}^\top \varepsilon\|_\infty \leq \eta_0 a_{*,\gamma}\right) \to 1$.

(b) $P\left(\|\varepsilon\|_\infty \leq \|\mathbf{V}\|_2 / \log^2 n\right) \to 1$.

(c) $P\left(n^{-1}\mathbf{V}^\top \varepsilon \geq n^{-1}\|\mathbf{V}\|_2^2 \rho_n\right) \to 1$.

We proceed in three steps with each step corresponding to one of the above claims. In the rest of the proof, we denote by $v_i$ the $i$-th entry of $\mathbf{V}$.

**Step 1: Establish the claim (a).**

Fix $\delta \geq 2$. For $1 \leq j \leq p$, define $L_{n,j} = \sum_{i=1}^n E|x_{i,j}\varepsilon_i|^{2+\delta} = nE|x_{1,j}\varepsilon_1|^{2+\delta}$ and $B_{n,j} = \sum_{i=1}^n E(x_{i,j}\varepsilon_i)^2 = nE(x_{1,j}\varepsilon_1)^2$. Since $x_{1,j}$ and $\varepsilon_1$ have bounded sub-Gaussian norms, $L_{n,j}/n$ is bounded above; since $E(x_{1,j}\varepsilon_1)^2$ is bounded away from zero and infinity, $B_{n,j}/n$ is bounded away from zero and infinity. Hence, there exists a constant $K_1 > 0$ such that for $1 \leq j \leq p$,

$$B_{n,j}/L_{n,j}^{1/(2+\delta)} \geq n^{\delta/(2+\delta)} K_1.$$

Let $a_n = \Phi^{-1}(1 - p^{-1}n^{-1})$. By Theorem 7.4 of (author?) [49], there exists an absolute constant $A > 0$ such that for $1 \leq j \leq p$,

$$P\left(\frac{\sum_{i=1}^n x_{i,j}\varepsilon_i}{\sqrt{\sum_{i=1}^n x_{i,j}^2 \varepsilon_i^2}} \geq a_n\right) \leq [1 - \Phi(a_n)]\left[1 + A\left(\frac{1 + a_n}{n^{\delta/(2+\delta)} K_1}\right)^{2+\delta}\right]$$

and

$$P\left(\frac{-\sum_{i=1}^n x_{i,j}\varepsilon_i}{\sqrt{\sum_{i=1}^n x_{i,j}^2 \varepsilon_i^2}} \geq a_n\right) \leq \Phi(-a_n)\left[1 + A\left(\frac{1 + a_n}{n^{\delta/(2+\delta)} K_1}\right)^{2+\delta}\right].$$

Since $\Phi(-a_n) = 1 - \Phi(a_n) = p^{-1}n^{-1}$ and $a_n = \Phi^{-1}(1 - p^{-1}n^{-1}) \leq \sqrt{2\log(pn)}$ (due to Lemma 1), the above two displays imply that, for $1 \leq j \leq p$,

$$P\left(\frac{|\sum_{i=1}^n x_{i,j}\varepsilon_i|}{\sqrt{\sum_{i=1}^n x_{i,j}^2 \varepsilon_i^2}} \geq a_n\right) \leq 2p^{-1}n^{-1}\left[1 + A\left(\frac{1 + \sqrt{2\log(pn)}}{n^{\delta/(2+\delta)} K_1}\right)^{2+\delta}\right].$$

By the union bound, we have that

$$P\left(\max_{1\leq j\leq p} \frac{|\sum_{i=1}^n x_{i,j}\varepsilon_i|}{\sqrt{\sum_{i=1}^n x_{i,j}^2 \varepsilon_i^2}} \geq a_n\right) \leq 2n^{-1}\left[1 + A\left(\frac{1 + \sqrt{2\log(pn)}}{n^{\delta/(2+\delta)} K_1}\right)^{2+\delta}\right] \stackrel{(i)}{=} o(1), \qquad (B.7)$$

where $(i)$ holds by $\log p = o(n)$ and $\delta \geq 2$. Since $\eta_0 = 1.1 n^{-1/2} a_n$, we have that

$$P\left(\max_{1\leq j\leq p} \frac{|n^{-1/2}\sum_{i=1}^n x_{i,j}\varepsilon_i|}{\sqrt{\sum_{i=1}^n x_{i,j}^2 \varepsilon_i^2}} \geq \eta_0\right) = o(1).$$

Claim (a) follows by the fact that

$$\frac{n^{-1}\|\mathbf{X}^\top \boldsymbol{\varepsilon}\|_\infty}{a_{*,\gamma}} = \frac{\max_{1\leq j\leq p}|n^{-1/2}\sum_{i=1}^n x_{i,j}\varepsilon_i|}{\max_{1\leq j\leq p}\sqrt{\sum_{i=1}^n x_{i,j}^2 \varepsilon_i^2}} \leq \max_{1\leq j\leq p} \frac{|n^{-1/2}\sum_{i=1}^n x_{i,j}\varepsilon_i|}{\sqrt{\sum_{i=1}^n x_{i,j}^2 \varepsilon_i^2}}.$$

**Step 2: Establish the claim (b).**

By the law of large numbers,

$$n^{-1}\|\mathbf{V}\|_2^2 = o_P(1) + Ev_1^2 = o_P(1) + E(\mathbf{x}_1^\top \boldsymbol{\gamma}^*)^2 + E\varepsilon_1^2 \geq o_P(1) + E\varepsilon_1^2.$$

Since $E\varepsilon_1^2$ is bounded away from zero, there exists a constant $M > 0$ such that $P(\|\mathbf{V}\|_2/\sqrt{n} \geq M) \to 1$. On the other hand, $\varepsilon_i$ has bounded sub-Gaussian norm and thus by the union bound, $\|\boldsymbol{\varepsilon}\|_\infty = O_P(\sqrt{\log n})$. Since $\sqrt{n}/\log^2 n \gg \sqrt{\log n}$, claim (b) follows.

**Step 3: Establish the claim (c).**

By the law of large numbers,
$$n^{-1}\mathbf{V}^\top \boldsymbol{\varepsilon} = o_P(1) + Ev_1\varepsilon_1 = o_P(1) + E\varepsilon_1^2.$$

Again, since $E\varepsilon_1^2$ is bounded away from zero, there exists a constant $M' > 0$ such that $P(n^{-1}\mathbf{V}^\top \boldsymbol{\varepsilon} \geq M') \to 1$. On the other hand, due to $n^{-1}\|\mathbf{V}\|_2^2 = Ev_1^2 + o_P(1)$, $\rho_n = o(1)$ and $Ev_1^2 = O(1)$, we have that $n^{-1}\|\mathbf{V}\|_2^2 \rho_n = o_P(1)$. Claim (c) follows.

We have showed the claims (a)-(c). The proof is complete. $\square$

**Lemma 3.** *Let Condition 1 hold. Let $\widehat{\sigma}_\varepsilon = \|\mathbf{V} - \mathbf{X}\widehat{\boldsymbol{\gamma}}\|_2/\sqrt{n}$. Suppose that $H_0$ in (1.2) holds. Then, with probability tending to one, we have that*
*(1) $\|n^{-1}\mathbf{X}^\top(\mathbf{V} - \mathbf{X}\widehat{\boldsymbol{\gamma}})\|_\infty / \widehat{\sigma}_\varepsilon \leq 2\|\mathbf{X}\|_\infty \eta_0 \rho_n^{-1}$ and*
*(2) $\|\mathbf{V} - \mathbf{X}\widehat{\boldsymbol{\gamma}}\|_\infty / \widehat{\sigma}_\varepsilon \leq \sqrt{n}/(\rho_n \log^2 n)$.*

*Proof of Lemma 3.* We define $a_{*,\gamma}^2 = \max_{1 \leq j \leq p} n^{-1} \sum_{i=1}^n x_{i,j}^2 \varepsilon_i^2$ and $a_{0,\gamma} = \|\mathbf{X}\|_\infty \|\mathbf{V}\|_2/\sqrt{n}$. We first show that $P(2a_{0,\gamma} \geq a_{*,\gamma}) \to 1$. By the law of large numbers,
$$n^{-1}\sum_{i=1}^n (4v_i^2 - \varepsilon_i^2) = E(4v_i^2 - \varepsilon_i^2) + o_P(1) = 3\sigma_\varepsilon^2 + 4E(\mathbf{x}_i^\top \boldsymbol{\gamma}^*)^2 + o_P(1), \tag{B.8}$$

where $\sigma_\varepsilon^2 = E\varepsilon_i^2$. Hence,

$$\begin{aligned}
P\left(4a_{0,\gamma}^2 \leq a_{*,\gamma}^2\right) &= P\left(4n^{-1}\|\mathbf{X}\|_\infty^2 \|\mathbf{V}\|_2^2 \leq \max_{1 \leq j \leq p} n^{-1}\sum_{i=1}^n x_{i,j}^2 \varepsilon_i^2\right) \\
&= P\left(4n^{-1}\|\mathbf{V}\|_2^2 \leq \max_{1 \leq j \leq p} n^{-1}\sum_{i=1}^n \|\mathbf{X}\|_\infty^{-2} x_{i,j}^2 \varepsilon_i^2\right) \\
&\leq P\left(4n^{-1}\|\mathbf{V}\|_2^2 \leq n^{-1}\sum_{i=1}^n \varepsilon_i^2\right) \\
&\overset{(i)}{\leq} P\left(3\sigma_\varepsilon^2 + 4E(\mathbf{x}_i^\top \boldsymbol{\gamma}^*)^2 + o_P(1) \leq 0\right) \overset{(ii)}{=} o(1),
\end{aligned}$$

where $(i)$ follows by (B.8) and $(ii)$ follows by the fact that $\sigma_\varepsilon$ is bounded away from zero. In other words, $P(2a_{0,\gamma} \geq a_{*,\gamma}) \to 1$.

Notice that the feasible set of the optimization problem (2.1) is increasing in $a$. By Lemma 2, $\boldsymbol{\gamma}^*$ lies in the feasibility set of the optimization problem (2.1) for $a = a_{*,\gamma}$ with probability approaching one. Hence, $P(2a_{0,\gamma} \geq a_{*,\gamma}) \to 1$ implies
$$P(\mathcal{A}) \to 1,$$
where $\mathcal{A}$ denotes the event that $\boldsymbol{\gamma}^*$ lies in the feasibility set of the optimization problem (2.1) for $a = 2a_{0,\gamma}$. Observe that on the event $\mathcal{A}$, $\widehat{\boldsymbol{\gamma}}$ is well defined.

By the last constraint in (2.1), we have that on the event $\mathcal{A}$,
$$n^{-1/2}\|\mathbf{V}\|_2 \widehat{\sigma}_\varepsilon = n^{-1}\|\mathbf{V}\|_2 \|\mathbf{V} - \mathbf{X}\widehat{\boldsymbol{\gamma}}\|_2 \geq n^{-1}\mathbf{V}^\top(\mathbf{V} - \mathbf{X}\widehat{\boldsymbol{\gamma}}) \geq \rho_n n^{-1}\|\mathbf{V}\|_2^2.$$

Hence, on the event $\mathcal{A}$,
$$\|\mathbf{V}\|_2 \leq \sqrt{n}\rho_n^{-1}\widehat{\sigma}_\varepsilon. \tag{B.9}$$

Define the event $\mathcal{M} = \{\mathcal{S}_\gamma = \emptyset\}$. On the event $\mathcal{M} \bigcap \mathcal{A}$, $\widehat{\gamma} = \widetilde{\gamma}(2a_{0,\gamma})$ and thus

$$\|n^{-1}\mathbf{X}^\top(\mathbf{V} - \mathbf{X}\widehat{\gamma})\|_\infty \leq 2\eta_0 a_{0,\gamma} = 2\eta_0 \|\mathbf{X}\|_\infty \|\mathbf{V}\|_2/\sqrt{n} \stackrel{(i)}{\leq} 2\eta_0 \|\mathbf{X}\|_\infty \rho_n^{-1} \widehat{\sigma}_\varepsilon, \tag{B.10}$$

where $(i)$ follows by (B.9).

On the event $\mathcal{M}^c \bigcap \mathcal{A}$, $\mathcal{S}_\gamma \neq \emptyset$, $\widehat{\gamma} = \widetilde{\gamma}(\widehat{a}_\gamma)$ for some $\widehat{\sigma}_\gamma \in \mathcal{S}_\gamma$ and therefore,

$$\frac{1}{2}\widehat{a}_\gamma \stackrel{(i)}{\leq} \sqrt{\max_{1 \leq j \leq p} n^{-1} \sum_{i=1}^n x_{i,j}^2 (v_i - \mathbf{x}_i^\top \widehat{\gamma})^2}$$
$$\leq \|\mathbf{X}\|_\infty \sqrt{n^{-1} \sum_{i=1}^n (v_i - \mathbf{x}_i^\top \widehat{\gamma})^2} = \|\mathbf{X}\|_\infty \widehat{\sigma}_\varepsilon, \tag{B.11}$$

where $(i)$ follows by the definition of $\mathcal{S}_\gamma$ in (2.2). Notice that on the event $\mathcal{M}^c \bigcap \mathcal{A}$,

$$\|n^{-1}\mathbf{X}^\top(\mathbf{V} - \mathbf{X}\widehat{\gamma})\|_\infty \stackrel{(i)}{\leq} \eta_0 \widehat{a}_\gamma \stackrel{(ii)}{\leq} 2\|\mathbf{X}\|_\infty \eta_0 \widehat{\sigma}_\varepsilon \tag{B.12}$$

where $(i)$ follows by the first constraint in (2.1) and the fact that $\widehat{\gamma} = \widetilde{\gamma}(\widehat{a}_\gamma)$ and $(ii)$ follows by (B.11). Since $P(\mathcal{A}) \to 1$ and $\rho_n \leq 1$, part (1) follows by (B.10) and (B.12).

In regards to part (2), notice that on the event $\mathcal{A}$,

$$\|\mathbf{V} - \mathbf{X}\widehat{\gamma}\|_\infty \stackrel{(i)}{\leq} \|\mathbf{V}\|_2/\log^2 n \stackrel{(ii)}{\leq} \frac{\sqrt{n}\widehat{\sigma}_\varepsilon}{\rho_n \log^2 n},$$

where $(i)$ follows by the second constraint in (2.1) and $(ii)$ follows by (B.9). Part (2) follows. $\square$

**Lemma 4.** *Let Condition 1 hold. Then, with probability tending to one, $\boldsymbol{\theta}^*$ lies in the feasible set of the optimization problem (2.4) for $a = a_{*,\theta}$, where $a_{*,\theta}^2 = \max_{1 \leq j \leq p} n^{-1} \sum_{i=1}^n x_{i,j}^2 u_i^2$.*

*Proof of Lemma 4.* The argument is identical to the proof of Lemma 2, except that $\mathbf{V}$, $\boldsymbol{\gamma}^*$, $a_{*,\gamma}$ and $\{\varepsilon_i\}_{i=1}^n$ are replaced by $\mathbf{Z}$, $\boldsymbol{\theta}^*$, $a_{*,\theta}$ and $\{u_i\}_{i=1}^n$, respectively. Since $\mathbf{Z} = \mathbf{X}\boldsymbol{\theta}^* + \mathbf{u}$ holds regardless of whether or not $H_0$ holds, the same reasoning in the proof of Lemma 2 applies. $\square$

**Lemma 5.** *Let Condition 1 hold. If $s/p \to 0$ and $s \log p = o(n)$, Then there exists a constant $K > 0$ such that*

$$P\left(\min_{A \subset \{1, \cdots, p\}, \, |A| \leq s} \min_{\|v_{A^c}\|_1 \leq \|v_A\|_1} \frac{\|n^{-1/2}\mathbf{X}v\|_2}{\|v_J\|_2} > K\right) \to 1.$$

*Proof of Lemma 5.* We invoke Theorem 6 of **(author?)** [58] with

$(\mathbf{A}, q, p, s_0, k_0, \delta) = (\boldsymbol{\Sigma}_X^{1/2}, p, p, s, 1, 1/2)$. Clearly, RE($s, 3, \mathbf{A}$) condition holds; see Definition 1 in **(author?)** [58].

Let $\boldsymbol{\Psi} = \mathbf{X}\boldsymbol{\Sigma}_X^{-1/2}$. Notice that rows of $\boldsymbol{\Psi}$ are independent vectors with mean zero and variance $\mathbb{I}_p$ and hence are isotropic random vectors; see Definition 5 of **(author?)** [58]. By Condition 1, rows of $\boldsymbol{\Psi}$ are also $\psi_2$ vectors for some constant $\alpha > 0$.

By the well-behaved eigenvalues of $\boldsymbol{\Sigma}_X$ in Condition 1, we have that $m$ defined in the statement of Theorem 6 of **(author?)** [58] satisfies $m \leq \min\{Cs, p\}$ for some constant $C > 0$. Since $s/p \to 0$ and $s \log p = o(n)$, we have that, for large $n$,

$$n \geq \frac{2000 m \alpha^4}{\delta^2} \log\left(\frac{60ep}{m\delta}\right).$$

By the well-behaved eigenvalues of $\boldsymbol{\Sigma}_X$, $K(s, 1, \mathbf{A})$ defined in **(author?)** [58] is bounded below by a constant $C_0 > 0$. Since $\mathbf{X} = \boldsymbol{\Psi}\mathbf{A}$, it follows, by Theorem 6 of therein (in particular Equation (14) therein), that for large $n$,

$$P\left(\min_{A \subset \{1, \cdots, p\}, \ |A| \leq s} \min_{\|v_{A^c}\|_1 \leq \|v_A\|_1} \frac{\|n^{-1/2}\mathbf{X}v\|_2}{\|v_J\|_2} > (1-\delta)/C_0\right) \geq 1 - 2\exp(-\delta^2 n / 2000\alpha^4).$$

The proof is complete. $\square$

**Lemma 6.** *Consider vectors $\mathbf{H} \in \mathbb{R}^n$ and $\mathbf{b}, \widehat{\mathbf{b}} \in \mathbb{R}^p$ and a matrix $\mathbf{M} \in \mathbb{R}^{n \times p}$. Suppose that for some $\sigma, \eta > 0$, $\|n^{-1}\mathbf{M}^\top(\mathbf{H} - \mathbf{Mb})\|_\infty \leq \eta\sigma$ and $\|n^{-1}\mathbf{M}^\top(\mathbf{H} - \mathbf{M}\widehat{\mathbf{b}})\|_\infty \leq \eta\sigma$. Assume that*

$$\min_{A \subset \{1, \cdots, p\}, \ |A| \leq s} \min_{\|v_{A^c}\|_1 \leq \|v_A\|_1} \frac{\|n^{-1/2}\mathbf{M}v\|_2}{\|v_J\|_2} > K \tag{B.13}$$

*for some $K > 0$. If $\|\mathbf{b}\|_0 \leq s$ and $\|\widehat{\mathbf{b}}\|_1 \leq \|\mathbf{b}\|_0$, then*

$$\|\widehat{\mathbf{b}} - \mathbf{b}\|_1 \leq 4K^{-2}\eta\sigma s$$

*and*

$$\left|\sqrt{\max_{1 \leq j \leq p} n^{-1}\sum_{i=1}^n M_{i,j}^2 (H_i - M_i^\top \widehat{\mathbf{b}})^2} - \sqrt{\max_{1 \leq j \leq p} n^{-1}\sum_{i=1}^n M_{i,j}^2 (H_i - M_i^\top \mathbf{b})^2}\right|$$
$$\leq 4\|\mathbf{M}\|_\infty \sigma K^{-1}\eta\sqrt{s},$$

*where $M_{i,j}$, $M_i^\top$ and $H_i$ denote the $(i, j)$ entry of $\mathbf{M}$, the $i$-th row of $\mathbf{M}$ and the $i$-th entry of $\mathbf{H}$, respectively and $\|\mathbf{M}\|_\infty = \max_{1 \leq j \leq p, \ 1 \leq i \leq n} |M_{i,j}|$.*

*Proof of Lemma 6.* The argument closely follows the proof of Theorem 7.1 of **(author?)** [6]. Define $J = \text{support}(b)$ and $\boldsymbol{\Delta} = \widehat{\mathbf{b}} - \mathbf{b}$. Since $\|\widehat{\mathbf{b}}\|_1 \leq \|\mathbf{b}\|_0$, it follows that $\|\widehat{\mathbf{b}}_J\|_1 + \|\boldsymbol{\Delta}_{J^c}\|_1 = \|\widehat{\mathbf{b}}_J\|_1 + \|\widehat{\mathbf{b}}_{J^c}\|_1 \leq \|\mathbf{b}_J\|_1$ and thus $\|\boldsymbol{\Delta}_{J^c}\|_1 \leq \|\boldsymbol{\Delta}_J\|_1$, implying that

$$\|\boldsymbol{\Delta}_J\|_1 \leq \|\boldsymbol{\Delta}\|_1 \leq 2\|\boldsymbol{\Delta}_J\|_1. \tag{B.14}$$

By the triangular inequality, we have

$$\|n^{-1}\mathbf{M}^\top \mathbf{M}\boldsymbol{\Delta}\|_\infty \leq \|n^{-1}\mathbf{M}^\top(\mathbf{H} - \mathbf{Mb})\|_\infty + \|n^{-1}\mathbf{M}^\top(\mathbf{H} - \mathbf{M}\widehat{\mathbf{b}})\|_\infty \leq 2\eta\sigma. \tag{B.15}$$

Therefore,

$$n^{-1}\boldsymbol{\Delta}^\top \mathbf{M}^\top \mathbf{M}\boldsymbol{\Delta} \overset{(i)}{\leq} \|\boldsymbol{\Delta}\|_1 \|n^{-1}\mathbf{M}^\top \mathbf{M}\boldsymbol{\Delta}\|_\infty \overset{(ii)}{\leq} 4\eta\sigma \|\boldsymbol{\Delta}_J\|_1 \overset{(iii)}{\leq} 4\eta\sigma\sqrt{s}\|\boldsymbol{\Delta}_J\|_2, \tag{B.16}$$

where $(i)$ follows by Holder's inequality, $(ii)$ follows by (B.14) and (B.15) and $(iii)$ follows by Holder's inequality and $|J| \leq s$. Since $\|\boldsymbol{\Delta}_{J^c}\|_1 \leq \|\boldsymbol{\Delta}_J\|_1$ and $|J| \leq s$, we have that $n^{-1}\boldsymbol{\Delta}^\top \mathbf{M}^\top \mathbf{M}\boldsymbol{\Delta} \geq K^2\|\boldsymbol{\Delta}_J\|_2^2$ (due to the assumption in (B.13)). This and the above display imply that

$$\|\boldsymbol{\Delta}_J\|_2 \leq 4K^{-2}\eta\sigma\sqrt{s}. \tag{B.17}$$

The first claim follows by Holder's inequality: $\|\boldsymbol{\Delta}_J\|_1 \leq \sqrt{s}\|\boldsymbol{\Delta}_J\|_2 \leq 4K^{-2}\eta\sigma s$.

To show the second claim, we define $\|\cdot\|_{M,j}$ on $\mathbb{R}^n$ by $\|a\|_{M,j} = \sqrt{n^{-1}\sum_{i=1}^n M_{i,j}^2 a_i^2}$ for $a = (a_1, ..., a_n)^\top$. For any $a, b \in \mathbb{R}^n$, we define $a_{M,j} = (M_{1,j}a_1, ..., M_{n,j}a_n)^\top \in \mathbb{R}^n$ and $b_{M,j} = (M_{1,j}b_1, ..., M_{n,j}b_n)^\top \in \mathbb{R}^n$ and notice that

$$\|a + b\|_{M,j} = \|a_{M,j} + b_{M,j}\|_2 \overset{(i)}{\leq} \|a_{M,j}\|_2 + \|b_{M,j}\|_2 = \|a\|_{M,j} + \|b\|_{M,j},$$

where $(i)$ follows by the triangular inequality for the Euclidean norm in $\mathbb{R}^n$. Hence, $\|\cdot\|_{M,j}$ is a semi-norm for any $1 \leq j \leq p$. The semi-norm property of $\|\cdot\|_{M,j}$ for all $1 \leq j \leq p$ implies that

$$\max_{1 \leq j \leq p} \left| \|\mathbf{H} - \mathbf{M}\widehat{\mathbf{b}}\|_{M,j} - \|\mathbf{H} - \mathbf{M}\mathbf{b}\|_{M,j} \right| \leq \max_{1 \leq j \leq p} \|\mathbf{M}\boldsymbol{\Delta}\|_{M,j} = \sqrt{n^{-1} \sum_{i=1}^n M_{i,j}^2 (M_i^\top \boldsymbol{\Delta})^2}$$

$$\leq \|\mathbf{M}\|_\infty \sqrt{n^{-1} \sum_{i=1}^n (M_i^\top \boldsymbol{\Delta})^2}$$

$$= \|\mathbf{M}\|_\infty \sqrt{n^{-1} \boldsymbol{\Delta}^\top \mathbf{M}^\top \mathbf{M} \boldsymbol{\Delta}}$$

$$\overset{(i)}{\leq} 4\|\mathbf{M}\|_\infty K^{-1} \eta \sigma \sqrt{s},$$

where $(i)$ follows by (B.16) and (B.17). The second claim follows by observing

$$\left| \max_{1 \leq j \leq p} \|\mathbf{H} - \mathbf{M}\widehat{\mathbf{b}}\|_{M,j} - \max_{1 \leq j \leq p} \|\mathbf{H} - \mathbf{M}\mathbf{b}\|_{M,j} \right|$$

$$\leq \max_{1 \leq j \leq p} \left| \|\mathbf{H} - \mathbf{M}\widehat{\mathbf{b}}\|_{M,j} - \|\mathbf{H} - \mathbf{M}\mathbf{b}\|_{M,j} \right|.$$

The proof is complete. $\square$

**Lemma 7.** *Let Condition 1 hold. Then $\|\widehat{\boldsymbol{\theta}} - \boldsymbol{\theta}^*\|_1 = O_P\left(n^{-1/2} s_\theta \log(p \vee n)\right)$ and $\|\mathbf{X}^\top(\mathbf{Z} - \mathbf{X}\widehat{\boldsymbol{\theta}})\|_\infty = O_P(\sqrt{n} \log(p \vee n))$.*

*Proof of Lemma 7.* We define $a_{*,\theta}^2 = \max_{1 \leq j \leq p} n^{-1} \sum_{i=1}^n x_{i,j}^2 u_i^2$ and the events

$$\mathcal{A}_1 = \{\boldsymbol{\theta}^* \text{ lies in the feasible set of the optimization problem (2.4) for } a = a_{*,\theta}.\}$$

$$\mathcal{A}_2 = \left\{ \min_{A \subset \{1,\cdots,p\},\, |A| \leq s} \min_{\|v_{A^c}\|_1 \leq \|v_A\|_1} \frac{\|n^{-1/2} \mathbf{X} v\|_2}{\|v_J\|_2} > K \right\}$$

$$\mathcal{A}_3 = \{a_{*,\theta} \in \mathcal{S}_\theta\}$$

$$\mathcal{A}_4 = \{\widehat{a}_\theta \leq 3 a_{*,\theta}\},$$

where $K > 0$ is the constant defined in Lemma 5. By Lemmas 4 and 5,

$$P\left(\mathcal{A}_1 \bigcap \mathcal{A}_2\right) \to 1. \tag{B.18}$$

We proceed in three steps. We first show that $P(\mathcal{A}_3) \to 1$ and then that $P(\mathcal{A}_4) \to 1$. Finally, we shall show the desired results.

**Step 1: Establish that $P(\mathcal{A}_3) \to 1$.** Notice that, on the event $\mathcal{A}_1 \bigcap \mathcal{A}_2$, $\|n^{-1}\mathbf{X}^\top(\mathbf{Z} - \mathbf{X}\widetilde{\boldsymbol{\theta}}(a_{*,\theta}))\|_\infty \leq \eta_0 a_{*,\theta}$, $\|n^{-1}\mathbf{X}^\top(\mathbf{Z} - \mathbf{X}\boldsymbol{\theta}^*)\|_\infty \leq \eta_0 a_{*,\theta}$ and $\|\widetilde{\boldsymbol{\theta}}(a_{*,\theta})\|_1 \leq \|\boldsymbol{\theta}^*\|_1$. Then we can apply Lemma 6 with

$$(\mathbf{H}, \mathbf{M}, \mathbf{b}, \widehat{\mathbf{b}}, \sigma, s, \eta) = (\mathbf{Z}, \mathbf{X}, \boldsymbol{\theta}^*, \widetilde{\boldsymbol{\theta}}(a_{*,\theta}), a_{*,\theta}, s_\theta, \eta_0)$$

and obtain that on the event $\mathcal{A}_1 \bigcap \mathcal{A}_2$,

$$\left| \sqrt{\max_{1 \leq j \leq p} n^{-1} \sum_{i=1}^n x_{i,j}^2 (z_i - \mathbf{x}_i^\top \widetilde{\boldsymbol{\theta}}(a_{*,\theta}))^2} - \sqrt{\max_{1 \leq j \leq p} n^{-1} \sum_{i=1}^n x_{i,j}^2 (z_i - \mathbf{x}_i^\top \boldsymbol{\theta}^*)^2} \right| \leq 4 a_{*,\theta} \|\mathbf{X}\|_\infty K^{-1} \eta_0 \sqrt{s_\theta}.$$

Notice that $a_{*,\theta}^2 = \max_{1\leq j\leq p} n^{-1} \sum_{i=1}^n x_{i,j}^2 (z_i - \mathbf{x}_i^\top \boldsymbol{\theta}^*)^2$. Thus, the above display implies that

$$\left| \frac{\sqrt{\max_{1\leq j\leq p} n^{-1} \sum_{i=1}^n x_{i,j}^2 (z_i - \mathbf{x}_i^\top \widetilde{\boldsymbol{\theta}}(a_{*,\theta}))^2}}{a_{*,\theta}} - 1 \right| \leq 4\|\mathbf{X}\|_\infty K^{-1} \eta_0 \sqrt{s_\theta} \stackrel{(i)}{=} o_P(1),$$

where $(i)$ holds by $\eta_0 \leq 1.1\sqrt{2n^{-1}\log(pn)}$ (due to Lemma 1), $\|\mathbf{X}\|_\infty = O_P(\sqrt{\log(pn)})$ (due to the sub-Gaussian property and the union bound) and the rate condition for $s_\theta$. Therefore,

$$P(\mathcal{A}_3) = P\left( \frac{1}{2} \leq \frac{\sqrt{\max_{1\leq j\leq p} n^{-1} \sum_{i=1}^n x_{i,j}^2 (z_i - \mathbf{x}_i^\top \widetilde{\boldsymbol{\theta}}(a_{*,\theta}))^2}}{a_{*,\theta}} \leq \frac{3}{2} \right) \to 1. \quad (B.19)$$

**Step 2: Establish that $P(\mathcal{A}_4) \to 1$.** On the event $\mathcal{A}_3$, $\mathcal{S}_\theta \neq \emptyset$ and thus $\widehat{\boldsymbol{\theta}} = \widetilde{\boldsymbol{\theta}}(\widehat{\sigma}_\theta)$ for some $\widehat{a}_\theta \geq a_{*,\theta}$ (since $\widehat{a}_\theta$ is the maximal element of $\mathcal{S}_\theta$). Notice that the feasible set of the optimization problem (2.4) is nondecreasing in $a$ and thus the mapping of $a \mapsto \|\widetilde{\boldsymbol{\theta}}(a)\|_1$ is non-increasing. Therefore, on the event $\mathcal{A}_1 \cap \mathcal{A}_2 \cap \mathcal{A}_3$, $\|\widetilde{\boldsymbol{\theta}}(\widehat{a}_\theta)\|_1 \leq \|\widetilde{\boldsymbol{\theta}}(a_{*,\theta})\|_1$. By $\|\widetilde{\boldsymbol{\theta}}(a_{*,\theta})\|_1 \leq \|\boldsymbol{\theta}^*\|_1$ (on the event $\mathcal{A}_1$), it follows that on the event $\mathcal{A}_1 \cap \mathcal{A}_2 \cap \mathcal{A}_3$,

$$\|\widetilde{\boldsymbol{\theta}}(\widehat{a}_\theta)\|_1 \leq \|\boldsymbol{\theta}^*\|_1, \quad \|n^{-1}\mathbf{X}^\top (\mathbf{V} - \mathbf{Z}\widetilde{\boldsymbol{\theta}}(\widehat{a}_\theta))\|_\infty \leq \eta_0 \widehat{a}_\theta$$

and $\|n^{-1}\mathbf{X}^\top(\mathbf{V} - \mathbf{Z}\boldsymbol{\theta}^*)\|_\infty \leq \eta_0 a_{*,\theta} \leq \eta_0 \widehat{a}_\theta.$

Hence, we can apply Lemma 6 with $(\mathbf{H}, \mathbf{M}, \mathbf{b}, \widehat{\mathbf{b}}, \sigma, s, \eta) = (\mathbf{Z}, \mathbf{X}, \boldsymbol{\theta}^*, \widetilde{\boldsymbol{\theta}}(\widehat{a}_\theta), \widehat{a}_\theta, s_\theta, \eta_0)$ and obtain that on the event $\mathcal{A}_1 \cap \mathcal{A}_2 \cap \mathcal{A}_3$,

$$\left| \sqrt{\max_{1\leq j\leq p} n^{-1} \sum_{i=1}^n x_{i,j}^2 (z_i - \mathbf{x}_i^\top \widetilde{\boldsymbol{\theta}}(\widehat{a}_\theta))^2} - a_{*,\theta} \right| / \widehat{a}_\theta$$

$$= \left| \sqrt{\max_{1\leq j\leq p} n^{-1} \sum_{i=1}^n x_{i,j}^2 (z_i - \mathbf{x}_i^\top \widetilde{\boldsymbol{\theta}}(\widehat{a}_\theta))^2} - \sqrt{\max_{1\leq j\leq p} n^{-1} \sum_{i=1}^n x_{i,j}^2 (z_i - \mathbf{x}_i^\top \boldsymbol{\theta}^*)^2} \right| / \widehat{a}_\theta$$

$$\leq 4\|\mathbf{X}\|_\infty K^{-1} \eta_0 \sqrt{s_\theta} = o_P(1).$$

In other words,

$$\sqrt{\max_{1\leq j\leq p} n^{-1} \sum_{i=1}^n x_{i,j}^2 (z_i - \mathbf{x}_i^\top \widetilde{\boldsymbol{\theta}}(\widehat{a}_\theta))^2} = a_{*,\theta} + o_P(1)\widehat{a}_\theta.$$

**Step 3: Establish the desired results.** Now notice that on the event $\mathcal{A}_1 \cap \mathcal{A}_2 \cap \mathcal{A}_3 \cap \mathcal{A}_4$,

$$\begin{cases} \|\widetilde{\boldsymbol{\theta}}(\widehat{a}_\theta)\|_1 \stackrel{(i)}{\leq} \|\boldsymbol{\theta}^*\|_1 \\ \|n^{-1}\mathbf{X}^\top(\mathbf{V} - \mathbf{Z}\widetilde{\boldsymbol{\theta}}(\widehat{a}_\theta))\|_\infty \stackrel{(ii)}{\leq} \eta_0 \widehat{a}_\theta \stackrel{(iii)}{\leq} 3\eta_0 a_{*,\theta} \\ \|n^{-1}\mathbf{X}^\top(\mathbf{V} - \mathbf{Z}\boldsymbol{\theta}^*)\|_\infty \stackrel{(iv)}{\leq} \eta_0 a_{*,\theta} \leq 3\eta_0 a_{*,\theta}. \end{cases}$$

where $(i)$ is proved in Step 2, $(ii)$ holds by the definition of $\mathcal{S}_\theta$ and $(iii)$ and $(iv)$ hold by the definitions of $\mathcal{A}_3$ and $\mathcal{A}_1$, respectively. Hence, we can apply Lemma 6 with $(\mathbf{H}, \mathbf{M}, \mathbf{b}, \widehat{\mathbf{b}}, \sigma, s, \eta) =$ $(\mathbf{Z}, \mathbf{X}, \boldsymbol{\theta}^*, \widetilde{\boldsymbol{\theta}}(\widehat{a}_\theta), 3a_{*,\theta}, s_\theta, \eta_0)$ and obtain that on the event $\mathcal{A}_1 \cap \mathcal{A}_2 \cap \mathcal{A}_3 \cap \mathcal{A}_4$,

$$\|\widetilde{\boldsymbol{\theta}}(\widehat{\sigma}_\theta) - \boldsymbol{\theta}^*\|_1 \leq 12 K^{-2} \eta_0 a_{*,\theta} s_\theta = 13.2 n^{-1/2} K^{-2} \Phi^{-1}(1 - n^{-1}p^{-1}) a_{*,\theta} s_\theta$$
$$\stackrel{(i)}{\leq} 13.2 n^{-1/2} K^{-2} \sqrt{2 \log(pn)} a_{*,\theta} s_\theta, \qquad (B.20)$$

where $(i)$ follows by Lemma 1. We notice that

$$a_{*,\theta}^2 = n^{-1} \max_{1 \leq j \leq p} \sum_{i=1}^n x_{i,j}^2 u_i^2 \leq \|\mathbf{X}\|_\infty^2 n^{-1} \sum_{i=1}^n u_i^2 \stackrel{(i)}{=} \|\mathbf{X}\|_\infty^2 [E(u_1^2) + o_P(1)]$$
$$\stackrel{(ii)}{=} O_P(\log(pn))[E(u_1^2) + o_P(1)]$$
$$= O_P(\log(pn)),$$

where $(i)$ follows by the law of large numbers and $(ii)$ follows by $\|\mathbf{X}\|_\infty = O_P(\sqrt{\log(pn)})$ (due to the sub-Gaussian property and the union bound). The first result follows by (B.20), the above display and $P(\mathcal{A}_1 \cap \mathcal{A}_2 \cap \mathcal{A}_3 \cap \mathcal{A}_4) \to 1$.

For the second result, we notice that on the event $\mathcal{A}_1 \cap \mathcal{A}_2 \cap \mathcal{A}_3 \cap \mathcal{A}_4$,

$$\|n^{-1} \mathbf{X}^\top (\mathbf{Z} - \mathbf{X} \widehat{\boldsymbol{\theta}})\|_\infty \stackrel{(i)}{\leq} \eta_0 \widehat{a}_\theta \stackrel{(ii)}{\leq} 3\eta_0 a_{*,\theta} \stackrel{(iii)}{=} O_P(n^{-1/2} \log(p \vee n))$$

where $(i)$ follows by the first constraint in (2.4) and the fact that $\widehat{\boldsymbol{\theta}} = \widetilde{\boldsymbol{\theta}}(\widehat{a}_\theta)$ on the event $\mathcal{A}_3$, $(ii)$ follows by the definition of $\mathcal{A}_4$ and $(iii)$ follows by $a_{*,\theta} = O_P(\sqrt{\log(pn)})$ (proved above) and $\eta_0 = O(\sqrt{n^{-1} \log(pn)})$ (due to Lemma 1). This proves the second result. $\square$

**Proof of Theorem 2.** Let $H_0$ hold. By Lemmas 2 and 7, we have that, with probability tending to one, the optimization problems (2.1) and (2.4) are feasible and thus the test statistic $T_n(\beta_0)$ is well defined. Recall that $\mathbf{V} = \mathbf{Y} - \mathbf{Z} \beta_0$. Since $H_0$ in (1.2) holds, we have

$$\mathbf{V} = \mathbf{X} \boldsymbol{\gamma}^* + \boldsymbol{\varepsilon}. \qquad (B.21)$$

Define $a_{*,\gamma}^2 = \max_{1 \leq j \leq p} \sum_{i=1}^n x_{i,j}^2 \varepsilon_i^2$. Let $\mathcal{A}$ denote the event that (i) $\boldsymbol{\gamma}^*$ lies in the feasible set of the optimization problem (2.1) for $a = a_{*,\gamma}$, (ii) $\|n^{-1} \mathbf{X}^\top (\mathbf{V} - \mathbf{X} \widehat{\boldsymbol{\gamma}})\|_\infty / \widehat{\sigma}_\varepsilon \leq 2 \|\mathbf{X}\|_\infty \eta_0 \rho_n^{-1}$ and (iii) $\|\mathbf{V} - \mathbf{X} \widehat{\boldsymbol{\gamma}}\|_\infty / \widehat{\sigma}_\varepsilon \leq \sqrt{n}/(\rho_n \log^2 n)$. By Lemmas 2 and 3,

$$P(\mathcal{A}) \to 1. \qquad (B.22)$$

Define $\xi_i = n^{-1/2} \widehat{\sigma}_\varepsilon^{-1} \varepsilon_i u_i \mathbf{1}\{\mathcal{A}\}$, $\widetilde{D}^2 = \sum_{i=1}^n \xi_i^2$ and $\widehat{D}^2 = \widehat{\sigma}_\varepsilon^{-2} n^{-1} \sum_{i=1}^n \widehat{\varepsilon}_i^2 \widehat{u}_i^2$, where $\widehat{\varepsilon}_i = v_i - \mathbf{x}_i^\top \widehat{\boldsymbol{\gamma}}$ and $\widehat{u}_i = z_i - \mathbf{x}_i^\top \widehat{\boldsymbol{\theta}}$. We define

$$\widetilde{T}_n = \frac{\widetilde{D}}{\widehat{D}} \times \underbrace{\frac{\sum_{i=1}^n \xi_i}{\widetilde{D}}}_{T_{n,1}} + \underbrace{\frac{n^{-1/2}(\mathbf{V} - \mathbf{X}\widehat{\boldsymbol{\gamma}})^\top \mathbf{X}(\boldsymbol{\theta}^* - \widehat{\boldsymbol{\theta}})}{\widehat{\sigma}_\varepsilon \widehat{D}}}_{T_{n,2}}.$$

By straight-forward computation, one can verify that on the event $\mathcal{A}$, $T_n(\beta_0) = \widetilde{T}_n$. By (B.22), it suffices to show that $\widetilde{T}_n \to^d N(0,1)$ under $H_0$. We do so in two steps. First, we show that $\widetilde{D}/\widehat{D} = 1 + o_P(1)$ and $T_{n,2} = o_P(1)$; second, we show $T_{n,1} \to^d N(0,1)$.

**Step 1: show that $\widetilde{D}/\widehat{D} = 1 + o_P(1)$ and $T_{n,2} = o_P(1)$.**

Notice that on the event $\mathcal{A}$,

$$\left|\widehat{D}^2 - \widetilde{D}^2\right| = \widehat{\sigma}_\varepsilon^{-2}\left|n^{-1}\sum_{i=1}^n \widehat{\varepsilon}_i^2(\widehat{u}_i^2 - u_i^2)\right| \leq \left(\max_{1\leq i\leq n}\left|\widehat{u}_i^2 - u_i^2\right|\right)\widehat{\sigma}_\varepsilon^{-2}n^{-1}\sum_{i=1}^n \widehat{\varepsilon}_i^2$$
$$= \|\mathbf{X}(\widehat{\boldsymbol{\theta}} - \boldsymbol{\theta}^*)\|_\infty^2$$
$$\leq \|\mathbf{X}\|_\infty^2\|\widehat{\boldsymbol{\theta}} - \boldsymbol{\theta}^*\|_1^2$$
$$\stackrel{(i)}{=} O_P(\log(pn))O_P\left(s_\theta^2 n^{-1}\log^2(p\vee n)\right)$$
$$= o_P(1), \tag{B.23}$$

where $(i)$ follows by Lemma 7 and $\|\mathbf{X}\|_\infty = O_P(\sqrt{\log(pn)})$ (due to the bounded sub-Gaussian norm of entries in $\mathbf{X}$ and the union bound). Let $\sigma_{u,i}^2 = E(u_i^2 \mid \mathbf{X}, \boldsymbol{\varepsilon})$. Observe that (B.22) implies

$$P\left(\left|\widetilde{D}^2 - n^{-1}\sum_{i=1}^n \sigma_{u,i}^2\right| = \widehat{\sigma}_\varepsilon^{-2}\left|n^{-1}\sum_{i=1}^n \widehat{\varepsilon}_i^2(u_i^2 - \sigma_{u,i}^2)\mathbf{1}\{\mathcal{A}\}\right|\right) \to 1. \tag{B.24}$$

Define $q_i = (u_i^2 - \sigma_{u,i}^2)\widehat{\varepsilon}_i^2\widehat{\sigma}_\varepsilon^{-2}\mathbf{1}\{\mathcal{A}\}$ and $d_n = \left(n^{-1}\widehat{\sigma}_\varepsilon^{-2}\max_{1\leq i\leq n}\widehat{\varepsilon}_i^2\right)\mathbf{1}\{\mathcal{A}\}$. By the definition of $\mathcal{A}$,

$$d_n = n^{-1}\|\mathbf{V} - \mathbf{X}\widehat{\boldsymbol{\gamma}}\|_\infty^2\widehat{\sigma}_\varepsilon^{-2}\mathbf{1}\{\mathcal{A}\} \stackrel{(i)}{\leq} \frac{1}{\rho_n^2 \log^4 n}, \tag{B.25}$$

where $(i)$ follows by the definition of $\mathcal{A}$.

Let $\mathcal{F}_{n,0}$ be the $\sigma$-algebra generated by $(\boldsymbol{\varepsilon}, \mathbf{X})$. By assumption, $\{u_i\}_{i=1}^n$ is independent across $i$ conditional on $\mathcal{F}_{n,0}$. Notice that $(\mathbf{V}, \mathbf{X})$ only depends on $(\boldsymbol{\varepsilon}, \mathbf{X})$ (due to (B.21)) and that $\widehat{\boldsymbol{\gamma}}$ and $\widehat{\sigma}_\varepsilon$ are computed using only $(\mathbf{V}, \mathbf{X})$. Therefore, $\{\widehat{\varepsilon}_i\}_{i=1}^n$, $\widehat{\sigma}_\varepsilon$ and $\mathbf{1}\{\mathcal{A}\}$ are $\mathcal{F}_{n,0}$-measurable.

Let $K > 0$ be a constant that upper bounds the sub-exponential norm of $u_i^2 - \sigma_{u,i}^2$ conditional on $\mathcal{F}_{n,0}$; such a constant $K$ exists because $u_i$ has bounded sub-Gaussian norm conditional on $\mathcal{F}_{n,0}$. Since $E(u_i^2 - \sigma_{u,i}^2 \mid \mathcal{F}_{n,0}) = 0$ and $\{u_i^2 - \sigma_{u,i}^2\}_{i=1}^n$ is independent across $i$ conditional on $\mathcal{F}_{n,0}$, we apply Proposition 5.16 of **(author?)** [67] to the conditional probability measure $P(\cdot \mid \mathcal{F}_{n,0})$ and obtain that there exists a universal constant $c > 0$ such that on the event $\mathcal{A}$, $\forall t > 0$,

$$P\left(\left|n^{-1}\sum_{i=1}^n q_i\right| > t \,\Big|\, \mathcal{F}_{n,0}\right) \leq 2\exp\left[-c\min\left\{\frac{t^2}{K^2 n^{-2}\sum_{i=1}^n \widehat{\varepsilon}_i^4\widehat{\sigma}_\varepsilon^{-4}}, \frac{t}{Kd_n}\right\}\right]$$
$$\stackrel{(i)}{\leq} 2\exp\left[-c\min\left\{\frac{t^2}{K^2 d_n^2}, \frac{t}{Kd_n}\right\}\right]$$
$$\stackrel{(ii)}{\leq} 2\exp\left[-c\min\left\{\frac{t^2 \rho_n^4 \log^8 n}{K^2}, \frac{t\rho_n^2 \log^4 n}{K}\right\}\right],$$

where $(i)$ follows by

$$\sum_{i=1}^n \widehat{\varepsilon}_i^4\widehat{\sigma}_\varepsilon^{-4} \leq \widehat{\sigma}_\varepsilon^{-2}\left(\max_{1\leq i\leq n}\widehat{\varepsilon}_i^2\right)\left(\widehat{\sigma}_\varepsilon^{-2}\sum_{i=1}^n \widehat{\varepsilon}_i^2\right) = n\widehat{\sigma}_\varepsilon^{-2}\max_{1\leq i\leq n}\widehat{\varepsilon}_i^2 = n^2 d_n$$

and $(ii)$ follows by (B.25). By (B.22) and $\rho_n \log^2 n \to \infty$, the above display implies that

$$P\left(\left|\widetilde{D}^2 - n^{-1}\sum_{i=1}^n \sigma_{u,i}^2\right| > t\right) \leq P(\mathcal{A}^c) + P\left(\left|n^{-1}\sum_{i=1}^n q_i\right| > t\right) = o(1) \quad \forall t > 0.$$

The above display implies that

$$\widetilde{D}^2 - n^{-1}\sum_{i=1}^n \sigma_{u,i}^2 = o_P(1). \tag{B.26}$$

It follows by (B.23) that

$$\widehat{D}^2 = n^{-1}\sum_{i=1}^{n}\sigma_{u,i}^2 + o_P(1). \tag{B.27}$$

Since $P(c_1 \leq n^{-1}\sum_{i=1}^{n}\sigma_{u,i}^2 \leq c_2) \to 1$ for some constants $c_1, c_2 > 0$, we have that

$$\frac{\widetilde{D}}{\widehat{D}} = \sqrt{\frac{n^{-1}\sum_{i=1}^{n}\sigma_{u,i}^2 + o_P(1)}{n^{-1}\sum_{i=1}^{n}\sigma_{u,i}^2 + o_P(1)}} = 1 + o_P(1).$$

We observe that

$$\begin{aligned}
|T_{n,2}| &\leq \frac{n^{-1/2}\|(\mathbf{V}-\mathbf{X}\widehat{\boldsymbol{\gamma}})^\top \mathbf{X}\|_\infty \|\widehat{\boldsymbol{\theta}}-\boldsymbol{\theta}^*\|_1}{\widehat{\sigma}_\varepsilon \widehat{D}} \\
&\overset{(i)}{\leq} \frac{2\sqrt{n}\|\mathbf{X}\|_\infty \eta_0 \rho_n^{-1} \|\widehat{\boldsymbol{\theta}}-\boldsymbol{\theta}^*\|_1}{\widehat{D}} \\
&\overset{(ii)}{=} \frac{2\sqrt{n}\|\mathbf{X}\|_\infty \eta_0 \rho_n^{-1} O_P\left(s_\theta n^{-1/2}\log(p\vee n)\right)}{\sqrt{n^{-1}\sum_{i=1}^{n}\sigma_{u,i}^2 + o_P(1)}} = o_P(1),
\end{aligned} \tag{B.28}$$

where $(i)$ follows by the definition of $\mathcal{A}$ and $(ii)$ follows by (B.27) and Lemma 7.

**Step 2: show that $T_{n,1} \to^d N(0,1)$.**

Define $\mathcal{F}_{n,i}$ as the $\sigma$-algebra generated by $(\mathbf{X}, \boldsymbol{\varepsilon}, u_1, ..., u_i)$. Since $E(u_{i+1} \mid \mathcal{F}_{n,i}) = 0$ (by assumption), we have that $\{(\xi_i, \mathcal{F}_{n,i})\}_{i=1}^{n}$ is a martingale difference array, i.e., $E(\xi_{i+1} \mid \mathcal{F}_{n,i}) = 0$. We invoke the martingale central limit theorem. By Theorem 3.4 of [25], it suffices to verify the following conditions.

(a) $\max_{1\leq i\leq n} |\xi_i| = o_P(1)$.
(b) $E\max_{1\leq i\leq n} \xi_i^2 = o(1)$.
(c) $\sum_{i=1}^{n}\xi_i^2 = n^{-1}\sum_{i=1}^{n}\sigma_{u,i}^2 + o_P(1)$.
(d) $\lim_{\delta\to 0}\liminf_{n\to\infty} P\left(n^{-1}\sum_{i=1}^{n}\sigma_{u,i}^2 > \delta\right) = 1$.

Notice that

$$\begin{aligned}
E\left(\max_{1\leq i\leq n}\xi_i^2\right) &\leq E\left[n^{-1}\widehat{\sigma}_\varepsilon^{-2}\|\mathbf{V}-\mathbf{X}\widehat{\boldsymbol{\gamma}}\|_\infty^2 \left(\max_{1\leq i\leq n} u_i^2\right) \mathbf{1}\{\mathcal{A}\}\right] \\
&\overset{(i)}{\leq} \frac{1}{\rho_n^2 \log^4 n} E\left(\max_{1\leq i\leq n} u_i^2\right) \\
&\overset{(ii)}{=} \frac{1}{\rho_n^2 \log^4 n} O(\log n) = o(1),
\end{aligned}$$

where $(i)$ follows by the definition of $\mathcal{A}$ and $(ii)$ follows by the uniformly bounded sub-exponential norm of $u_i^2$ and Corollary 2.6 of [9]. This proves claim (b). Notice that claim (a) follows by claim (b). By the definition of $\widetilde{D}$, claim (c) follows by (B.26). Claim (d) follows by the assumption that $P(c_1 \leq n^{-1}\sum_{i=1}^{n}\sigma_{u,i}^2 \leq c_2) \to 1$ for some constants $c_1, c_2 > 0$.

The proof is complete. □

## B.3 Proof of Corollary 5

***Proof of Corollary 5.*** Similar to the comment made in regards to the Corollary 2.1 of [63], the proof of Corollary 5 is exactly the same as in Theorem 2 once we notice the following.

Take an arbitrary sequence $(\beta_{0n}, \xi_n) \in \mathbb{R} \times \Xi(s_\theta)$. Notice that the properties of the test statistic $T_n(\beta_{0n})$ depend on $\mathbf{X}, \mathbf{Z}$ and $\mathbf{V}$ exclusively. Moreover observe that under $H_0$, $\mathbf{V} = \mathbf{X}\boldsymbol{\gamma} + \boldsymbol{\varepsilon}$.

Moreover, in the arguments in the proof of Lemmas 2-7 and Theorem 2, we use bounds that hold in finite samples with universal constants as well as law of large numbers and martingale central limit theorem for triangular arrays. Hence, Lemmas 2-7 and the arguments in the proof of Theorem 2 still go through when $\xi_n$ is a sequence. □

## B.4 Proof of Theorems 6 and 7

**Proof of Theorem 6.** It suffices to notice that Theorem 6 is a special case of Theorem 8 in which the decomposition now reads $\boldsymbol{\gamma}^* = \boldsymbol{\pi}^* + \boldsymbol{\mu}^*$ with $\boldsymbol{\pi}^* = \boldsymbol{\gamma}^*$ and $\boldsymbol{\mu}^* = 0$. The result then follows by Theorem 8. □

**Proof of Theorem 7.** It suffices to notice that Theorem 7 is a special case of Theorem 8 in which the decomposition now reads $\boldsymbol{\gamma}^* = \boldsymbol{\pi}^* + \boldsymbol{\mu}^*$ with $\boldsymbol{\pi}^* = 0$ and $\boldsymbol{\mu}^* = \boldsymbol{\gamma}^*$. The result then follows by Theorem 8. □

## B.5 Proof of Theorem 8

The following result is due to [7] and pointed out by [41].

**Lemma 8.** *Suppose that $W_1, ..., W_n$ are i.i.d random variables such that $EW_1 = 0$, $EW_1^2 \geq C_1$ and $P(|W_1| > t) \leq C_2 \exp(-C_3 t^\alpha)$ for any $t > 0$, where $C_1, C_2, C_3 > 0$ and $\alpha \in (0, 1)$ are constants. Then for any $z > 0$,*

$$P\left(\left|n^{-1}\sum_{i=1}^n W_i\right| > z\right) \leq n \exp\left(-M_1(nz)^\alpha\right) + \exp\left(-M_2(nz)^2\right),$$

*where $M_1, M_2 > 0$ are constants depending only on $C_1, C_2, C_3$ and $\alpha$.*

*Proof.* Notice that $|W_1|^\alpha$ has bounded sub-exponential norm. Hence, $E\exp(D_1|W_1|^\alpha) \leq D_2$ for some constants $D_1, D_2 > 0$. Hence, by Equation (1.4) of [41], we have that for any $z > 0$,

$$P\left(\left|\sum_{i=1}^n W_i\right| > nz\right) \leq n \exp\left(-D_3(nz)^\alpha\right) + \exp\left(-D_4(nz)^2\right).$$

The proof is complete. □

**Lemma 9.** *Suppose that $\{(w_{i,1}, \ldots, w_{i,p})\}_{i=1}^n$ is an i.i.d sequence of p-dimensional random vectors with $m_j = E(w_{1,j})$ and $\log p = o(\sqrt{n})$.*

*Suppose that $\min_{1 \leq j \leq p} Var(w_{i,j}) \geq 2K_1$ and for all $1 \leq j \leq p$, the sub-exponential norm of $w_{1,j}$ is upper bounded by $K_2$, where $K_1, K_2 > 0$ are constants. If $\max_{1 \leq j \leq p} |m_j| \leq D\Phi(1 - p^{-1}n^{-1})/\sqrt{n}$ with $D \leq 0.04\sqrt{K_1}$, then*

$$P\left(\max_{1 \leq j \leq p}\left|n^{-1}\sum_{i=1}^n w_{i,j}\right| \geq \eta_0 A\right) \to 0,$$

*where $A^2 = \max_{1 \leq j \leq p} n^{-1}\sum_{i=1}^n w_{i,j}^2$.*

*Proof.* Define $t = \Phi(1 - p^{-1}n^{-1})$ and $\xi_{i,j} = w_{i,j} - m_j$ as well as the following events

$$\mathcal{M}_1 = \left\{ \max_{1 \leq j \leq p} \frac{\sum_{i=1}^n \xi_{i,j}}{\sqrt{\sum_{i=1}^n \xi_{i,j}^2}} \leq t \right\}$$

$$\mathcal{M}_2 = \left\{ \min_{1 \leq j \leq p} n^{-1} \sum_{i=1}^n \xi_{i,j}^2 \geq K_1 \right\} \cap \left\{ \min_{1 \leq j \leq p} n^{-1} \sum_{i=1}^n w_{i,j}^2 \geq K_1 \right\}$$

$$\mathcal{M}_3 = \left\{ \max_{1 \leq j \leq p} \left| n^{-1} \sum_{i=1}^n \xi_{i,j} \right| \leq K_3 \sqrt{n^{-1} \log p} \right\}$$

$$\mathcal{M}_4 = \left\{ \max_{1 \leq j \leq p} \left( Dt + t \sqrt{n^{-1} \sum_{i=1}^n \xi_{i,j}^2} - 1.05 t \sqrt{n^{-1} \sum_{i=1}^n w_{i,j}^2} \right) \leq 0 \right\},$$

where $K_3 > 0$ is a constant to be chosen. The proof proceeds in two steps. We first prove that $\mathcal{M}_1$, $\mathcal{M}_2$, $\mathcal{M}_3$ and $\mathcal{M}_4$ occur with probability approaching one and then show the desired result.

**Step 1: show that $\mathcal{M}_1$, $\mathcal{M}_2$, $\mathcal{M}_3$ and $\mathcal{M}_4$ occur with probability approaching one.**

Fix any $\delta \geq 2$. Notice that $E\xi_{i,j}^2$ is bounded away from zero and $E|\xi_{i,j}|^{2+\delta}$ is bounded away from infinity. As in the proof of Lemma 2 (Step 1), Theorem 7.4 of [49] implies that there exist constants $A, C_1 > 0$ such that for any $1 \leq j \leq p$,

$$P\left( \frac{|\sum_{i=1}^n \xi_{i,j}|}{\sqrt{\sum_{i=1}^n \xi_{i,j}^2}} \geq t \right) \leq 2\Phi(-t) \left[ 1 + A \left( \frac{1+t}{n^{\delta/(2+\delta)} C_1} \right)^{2+\delta} \right].$$

By the union bound, we have that

$$P(\mathcal{M}_1) = P\left( \max_{1 \leq j \leq p} \frac{|\sum_{i=1}^n \xi_{i,j}|}{\sqrt{\sum_{i=1}^n \xi_{i,j}^2}} \geq t \right) \leq 2p\Phi(-t) \left[ 1 + A \left( \frac{1+t}{n^{\delta/(2+\delta)} C_1} \right)^{2+\delta} \right] \stackrel{(i)}{=} o(1),$$

where $(i)$ follows by the fact that $\Phi(-t) = 1 - \Phi(t) = p^{-1}n^{-1}$ and $t = \Phi^{-1}(1 - p^{-1}n^{-1}) \leq \sqrt{2 \log(pn)}$ (due to Lemma 1). By the bounded sub-exponential norm of $\xi_{i,j}$, there exist constants $C_2, C_3 > 0$ such that for any $1 \leq j \leq p$ and any $z > 0$, $P(\xi_{1,j}^2 > z) \leq C_1 \exp(-C_2 \sqrt{n})$. Hence,

$$P\left( n^{-1} \sum_{i=1}^n \xi_{i,j}^2 \leq K_1 \right) = P\left( n^{-1} \sum_{i=1}^n (\xi_{i,j}^2 - E\xi_{i,j}^2) \leq K_1 - E\xi_{i,j}^2 \right)$$

$$\leq P\left( n^{-1} \sum_{i=1}^n (\xi_{i,j}^2 - E\xi_{i,j}^2) \leq -K_1 \right)$$

$$\stackrel{(i)}{\leq} n \exp\left( -C_3 \sqrt{nK_1} \right) + \exp\left( -C_4 (nK_1)^2 \right)$$

for some constants $C_3, C_4 > 0$, where $(i)$ follows by Lemma 8. By the union bound and $\log p = o(\sqrt{n})$, we have that

$$P\left( \min_{1 \leq j \leq p} n^{-1} \sum_{i=1}^n \xi_{i,j}^2 \leq K_1 \right) \leq \sum_{j=1}^p P\left( n^{-1} \sum_{i=1}^n \xi_{i,j}^2 \leq K_1 \right)$$

$$\leq pn \exp\left( -C_3 \sqrt{nK_1} \right) + p \exp\left( -C_4 (nK_1)^2 \right) = o(1).$$

Similarly, the above display holds if $\xi_{i,j}$ is replaced by $w_{i,j}$. Hence,
$$P(\mathcal{M}_2) \to 1.$$

Moreover, notice that
$$P\left(\max_{1\leq j\leq p}\left|n^{-1}\sum_{i=1}^n \xi_{i,j}\right| \geq K_3\sqrt{n^{-1}\log p}\right) \leq \sum_{j=1}^p P\left(\left|n^{-1}\sum_{i=1}^n \xi_{i,j}\right| \geq K_3\sqrt{n^{-1}\log p}\right)$$
$$\stackrel{(i)}{\leq} 2p\exp\left[-c\min\left(\frac{K_3^2 \log p}{K_2^2}, \frac{K_3\sqrt{n\log p}}{K_2}\right)\right]$$

for some universal constant $c > 0$, where $(i)$ follows by Proposition 5.16 of [67]. Hence, for $K_3 = 2K_2/\sqrt{c}$,
$$P(\mathcal{M}_3) \to 1.$$

Recall the elementary inequality that for any $a, b \geq 0$, $a + b + 1/2 \geq \sqrt{a} + \sqrt{b}$ and thus
$$\left|\sqrt{a} - \sqrt{b}\right| = \frac{|a-b|}{\sqrt{a}+\sqrt{b}} \geq \frac{|a-b|}{a+b+1/2}. \tag{B.29}$$

Moreover, observe that
$$tn^{-1}\sqrt{\log p} = \Phi^{-1}(1 - n^{-1}p^{-1})n^{-1}\sqrt{\log p} \stackrel{(i)}{=} O\left(n^{-1}\sqrt{(\log p)(\log(pn))}\right) \stackrel{(ii)}{=} o(1), \tag{B.30}$$

where $(i)$ follows by Lemma 1 and $(ii)$ holds by $\log p = o(\sqrt{n})$.

Hence, on the event $\mathcal{M}_1 \cap \mathcal{M}_2 \cap \mathcal{M}_3$,

$$\sqrt{n^{-1}\sum_{i=1}^n w_{i,j}^2} \stackrel{(i)}{\geq} \sqrt{n^{-1}\sum_{i=1}^n \xi_{i,j}^2} - \frac{\left|n^{-1}\sum_{i=1}^n (w_{i,j}^2 - \xi_{i,j}^2)\right|}{1/2 + n^{-1}\sum_{i=1}^n (w_{i,j}^2 + \xi_{i,j}^2)}$$
$$= \sqrt{n^{-1}\sum_{i=1}^n \xi_{i,j}^2} - \frac{\left|n^{-1}\sum_{i=1}^n (2m_j\xi_{i,j} + m_j^2)\right|}{1/2 + n^{-1}\sum_{i=1}^n (2\xi_{i,j}^2 + 2m_j\xi_{i,j} + m_j^2)}$$
$$= \sqrt{n^{-1}\sum_{i=1}^n \xi_{i,j}^2} - \frac{\left|m_j^2 + 2m_j\left(n^{-1}\sum_{i=1}^n \xi_{i,j}\right)\right|}{1/2 + m_j^2 + 2\left(n^{-1}\sum_{i=1}^n \xi_{i,j}^2\right) + 2m_j\left(n^{-1}\sum_{i=1}^n \xi_{i,j}\right)}$$
$$\stackrel{(ii)}{\geq} \sqrt{n^{-1}\sum_{i=1}^n \xi_{i,j}^2} - \frac{m_j^2 + 2|m_j|K_3\sqrt{n^{-1}\log p}}{1/2 + m_j^2 + 2K_1 - 2|m_j|K_3\sqrt{n^{-1}\log p}}$$
$$\geq \sqrt{n^{-1}\sum_{i=1}^n \xi_{i,j}^2} - \frac{D^2t^2n^{-1} + 2Dtn^{-1}K_3\sqrt{\log p}}{1/2 + 2K_1 - 2Dtn^{-1}K_3\sqrt{\log p}}$$
$$\stackrel{(iii)}{\geq} \sqrt{n^{-1}\sum_{i=1}^n \xi_{i,j}^2} - \frac{D^2t^2n^{-1} + 2Dtn^{-1}K_3\sqrt{\log p}}{1/2 + K_1}, \tag{B.31}$$

where $(i)$ follows by the previous elementary inequality in (B.29), $(ii)$ follows by the definitions of $\mathcal{M}_2$ and $\mathcal{M}_3$, $(iii)$ follows by the fact that $2Dtn^{-1}K_3\sqrt{\log p} < K_1$ for large $n$ (due to (B.30)). Now we have that on the event $\mathcal{M}_1 \cap \mathcal{M}_2 \cap \mathcal{M}_3$,

$$\max_{1 \leq j \leq p} \left( Dt + t\sqrt{n^{-1}\sum_{i=1}^n \xi_{i,j}^2} - 1.05t\sqrt{n^{-1}\sum_{i=1}^n w_{i,j}^2} \right) / t$$

$$\stackrel{(i)}{\leq} \max_{1 \leq j \leq p} \left( Dt - 0.05t\sqrt{n^{-1}\sum_{i=1}^n w_{i,j}^2} + t\frac{D^2t^2n^{-1} + 2Dtn^{-1}K_3\sqrt{\log p}}{1/2 + K_1} \right) / t$$

$$\stackrel{(ii)}{\leq} D - 0.05\sqrt{K_1} + \frac{D^2t^2n^{-1} + 2Dtn^{-1}K_3\sqrt{\log p}}{1/2 + K_1}$$

$$\stackrel{(iii)}{=} D - 0.05\sqrt{K_1} + o(1).$$

where $(i)$ follows by (B.31), $(ii)$ follows by the definition of $\mathcal{M}_2$ and $(iii)$ follows by (B.30).
Since $D \leq 0.04\sqrt{K_1}$, the above display implies that

$$P(\mathcal{M}_4) = P\left( \max_{1 \leq j \leq p} \left( Dt + t\sqrt{n^{-1}\sum_{i=1}^n \xi_{i,j}^2} - 1.05t\sqrt{n^{-1}\sum_{i=1}^n w_{i,j}^2} \right) \leq 0 \right) \to 1.$$

**Step 2: prove the desired result.**
The desired result now follows by the following observation

$$P\left( \left| n^{-1}\sum_{i=1}^n w_{i,j} \right| \geq \eta_0 A \right)$$

$$= P\left( \max_{1 \leq j \leq p} \left| \sqrt{n}m_j + n^{-1/2}\sum_{i=1}^n \xi_{i,j} \right| \geq 1.1t \max_{1 \leq j \leq p}\sqrt{n^{-1}\sum_{i=1}^n w_{i,j}^2} \right)$$

$$\leq P\left( Dt + \max_{1 \leq j \leq p} \left| \sum_{i=1}^n \xi_{i,j} \right| \geq 1.1t \max_{1 \leq j \leq p}\sqrt{n^{-1}\sum_{i=1}^n w_{i,j}^2} \right)$$

$$\leq P\left( Dt + \max_{1 \leq j \leq p} \left| \sum_{i=1}^n \xi_{i,j} \right| \geq 1.1t \max_{1 \leq j \leq p}\sqrt{n^{-1}\sum_{i=1}^n w_{i,j}^2} \text{ and } \mathcal{M}_1 \right) + P(\mathcal{M}_1^c)$$

$$\leq P\left( Dt + \max_{1 \leq j \leq p} t\sqrt{n^{-1}\sum_{i=1}^n \xi_{i,j}^2} \geq 1.1t \max_{1 \leq j \leq p}\sqrt{n^{-1}\sum_{i=1}^n w_{i,j}^2} \right) + P(\mathcal{M}_1^c)$$

$$\leq P\left( Dt + \max_{1 \leq j \leq p} t\sqrt{n^{-1}\sum_{i=1}^n \xi_{i,j}^2} \geq 1.1t \max_{1 \leq j \leq p}\sqrt{n^{-1}\sum_{i=1}^n w_{i,j}^2} \text{ and } \mathcal{M}_4 \right) + P(\mathcal{M}_1^c) + P(\mathcal{M}_4^c)$$

$$\leq P\left( 1.05t \max_{1 \leq j \leq p}\sqrt{n^{-1}\sum_{i=1}^n w_{i,j}^2} \geq 1.1t \max_{1 \leq j \leq p}\sqrt{n^{-1}\sum_{i=1}^n w_{i,j}^2} \right) + P(\mathcal{M}_1^c) + P(\mathcal{M}_4^c)$$

$$= P(\mathcal{M}_1^c) + P(\mathcal{M}_4^c) \stackrel{(i)}{=} o(1),$$

where $(i)$ holds by Step 1. $\square$

**Lemma 10.** *Let the conditions in the statment of Theorem 8 hold. Suppose that $H_{1,h}$ in (3.4) holds. Define $\boldsymbol{\gamma}_{h,n} = \boldsymbol{\pi}^* + n^{-1/2}h\boldsymbol{\theta}^*$ and $\varepsilon_{(h),i} = \mathbf{x}_i^\top \boldsymbol{\mu}^* + n^{-1/2}hu_i + \varepsilon_i$. Then, with probability* *tending to one,* $\gamma_{h,n}$ *lies in the feasible set of the optimization problem* (2.1) *for* $a = a_{h,\gamma}$, *where* $a_{h,\gamma}^2 = n^{-1} \max_{1 \leq j \leq p} \sum_{i=1}^n x_{i,j}^2 \varepsilon_{(h),i}^2$.

*Moreover,* $\max_{1 \leq j \leq p} |n^{-1} \sum_{i=1}^n x_{i,j} \varepsilon_{(h),i}| = O_P(n^{-1/2} \log(p \vee n))$.

*Proof of Lemma 10.* Let $\mathbf{V} = \mathbf{Y} - \mathbf{Z}\beta_0$ and $\boldsymbol{\varepsilon}_{(h)} = (\varepsilon_{(h),1}, \ldots, \varepsilon_{(h),n})^\top$. In the rest of the proof, we denote by $v_i$ the $i$-th entry of $\mathbf{V}$. Notice that under $H_{1,h}$ in (3.4),

$$v_i = y_i - z_i \beta_0 = n^{-1/2} h z_i + \mathbf{x}_i^\top \boldsymbol{\gamma}^* + \varepsilon_i = n^{-1/2} h (\mathbf{x}_i^\top \boldsymbol{\theta}^* + u_i) + \mathbf{x}_i^\top \boldsymbol{\gamma}^* + \varepsilon_i$$
$$= \mathbf{x}_i^\top \boldsymbol{\gamma}_{h,n} + \varepsilon_{(h),i}.$$

Then for the first result, it suffices to verify the following claims:

(a) $P\left(\|n^{-1} \mathbf{X}^\top \boldsymbol{\varepsilon}_{(h)}\|_\infty \leq \eta_0 a_{*,\gamma}\right) \to 1.$

(b) $P\left(\|\boldsymbol{\varepsilon}_{(h)}\|_\infty \leq \|\mathbf{V}\|_2 / \log^2 n\right) \to 1.$

(c) $P\left(n^{-1} \mathbf{V}^\top \boldsymbol{\varepsilon}_{(h)} \geq n^{-1} \|\mathbf{V}\|_2^2 \rho_n\right) \to 1.$

We proceed in four steps corresponding to above three claims as well as the second result.

**Step 1: Establish claim (a).** We apply Lemma 9. Let $w_{i,j} = x_{i,j} \varepsilon_{(h),i}$ and notice that

$$Var(w_{1,j}) = Var[x_{1,j}(\mathbf{x}_1^\top \boldsymbol{\mu}^* + \varepsilon_1 + n^{-1/2} h u_1)]$$
$$= n^{-1} h^2 E u_1^2 + Var[x_{1,j}(\mathbf{x}_1^\top \boldsymbol{\mu}^* + \varepsilon_1)] \stackrel{(i)}{\geq} 2K_1 \quad \text{with } K_1 = \tau_2/2,$$

where $(i)$ holds by the assumption of Theorem 8. Then

$$\max_{1 \leq j \leq p} |E w_{1,j}| = \|\boldsymbol{\Sigma}_X \boldsymbol{\mu}^*\|_\infty \stackrel{(i)}{\leq} \frac{1}{25\sqrt{2}} \sqrt{\tau_2 n^{-1} \log p}$$
$$\stackrel{(ii)}{\leq} \frac{1}{25\sqrt{2}} \sqrt{\tau_2 n^{-1}} \Phi^{-1}(1 - p^{-1})$$
$$< \frac{1}{25\sqrt{2}} \sqrt{\tau_2 n^{-1}} \Phi^{-1}(1 - n^{-1} p^{-1})$$
$$= 0.04 \sqrt{K_1} \Phi^{-1}(1 - n^{-1} p^{-1}) / \sqrt{n},$$

where $(i)$ holds by the assumption of Theorem 8 and $(ii)$ follows by Lemma 1. Since both $x_{i,j}$ and $\varepsilon_{(h),i}$ have bounded sub-Gaussian norms, it follows that $w_{i,j}$ has a bounded sub-exponential norm. Therefore, claim (a) follows by Lemma 9.

**Step 2: Establish claim (b).** By the law of large numbers

$$n^{-1} \|\mathbf{V}\|_2^2 - E v_1^2 = n^{-1} \sum_{i=1}^n (v_i^2 - E v_i^2) = o_P(1). \tag{B.32}$$

Notice that $E v_1^2 = E(\mathbf{x}_1^\top (\boldsymbol{\gamma}_{h,n} + \boldsymbol{\mu}^*))^2 + n^{-1} h^2 E u_1^2 + E \varepsilon_1^2 \geq E \varepsilon_1^2$ is bounded away from zero. Hence, there exists a constant $M > 0$ such that $P(\|\mathbf{V}\|_2 \geq \sqrt{n} M) \to 1$. On the other hand, $\varepsilon_{(h),i}$ has a bounded sub-Gaussian norm and thus $\|\boldsymbol{\varepsilon}_{(h)}\|_\infty = O_P(\sqrt{\log n})$. Since $\sqrt{n}/\log^2 n \gg \sqrt{\log n}$, claim (b) follows.

**Step 3: Establish claim (c).** Notice that

$$n^{-1}\mathbf{V}^\top \boldsymbol{\varepsilon}_{(h)} = E[v_1\varepsilon_{(h),1}] + n^{-1}\sum_{i=1}^n (v_i\varepsilon_{(h),i} - E[v_i\varepsilon_{(h),i}])$$

$$\stackrel{(i)}{=} E[v_1\varepsilon_{(h),1}] + o_P(1)$$

$$= (\boldsymbol{\pi}^* + n^{-1/2}h\boldsymbol{\theta}^* + \boldsymbol{\mu}^*)^\top \boldsymbol{\Sigma}_X \boldsymbol{\mu}^* + E(\varepsilon_1^2) + n^{-1}h^2 E(u_1^2) + o_P(1)$$

$$\stackrel{(ii)}{=} (\boldsymbol{\pi}^* + \boldsymbol{\mu}^*)^\top \boldsymbol{\Sigma}_X \boldsymbol{\mu}^* + E(\varepsilon_1^2) + o_P(1),$$

where $(i)$ holds by the law of large numbers and $(ii)$ holds by the fact that $\|\boldsymbol{\theta}^*\|_2$, $\|\boldsymbol{\mu}^*\|_2$, the eigenvalues of $\boldsymbol{\Sigma}_X$ and $E(u_1^2)$ are bounded. By assumption, $(\boldsymbol{\pi}^* + \boldsymbol{\mu}^*)^\top \boldsymbol{\Sigma}_X \boldsymbol{\mu}^* + E(\varepsilon_1^2)$ is bounded below by a positive constant. Hence, we have that

$$P(n^{-1}\mathbf{V}^\top \boldsymbol{\varepsilon}_{(h)} \geq M') \to 1$$

for some constant $M' > 0$. By (B.32) and the boundedness of $E(v_1^2)$, we have that $n^{-1}\|\mathbf{V}\|_2^2 \rho_n = O_P(\rho_n) = o_P(1)$. Hence, claim (c) follows.

We have showed the claims (a)-(c), completing the proof for the first result.

**Step 4: Establish the second result.** By claim (a) above and $\eta_0 \leq 1.1\sqrt{2n^{-1}\log(pn)}$ (due to Lemma 1), it suffices to show that $a_{*,\gamma} = O_P(\sqrt{\log(pn)})$, which is obtained by the following observation

$$a_{*,\gamma}^2 = \max_{1 \leq j \leq p} n^{-1}\sum_{i=1}^n x_{i,j}^2 \varepsilon_{(h),i}^2 \leq \|\mathbf{X}\|_\infty^2 n^{-1}\sum_{i=1}^n \varepsilon_{(h),i}^2 \stackrel{(i)}{=} \|\mathbf{X}\|_\infty^2 (E\varepsilon_{(h),1}^2 + o_P(1))$$

$$\stackrel{(ii)}{=} O_P(\log(pn))O_P(1),$$

where $(i)$ follows by the law of large numbers and $(ii)$ follows by the sub-Gaussian properties of entries in $\mathbf{X}$ and the union bound. We have proved the second result. $\square$

**Lemma 11.** *Let the conditions in the statment of Theorem 8 hold. Suppose that $H_{1,h}$ in (3.4) holds. Then $\|\widehat{\boldsymbol{\gamma}} - \boldsymbol{\gamma}_{h,n}\|_1 = O_P\left(n^{-1/2}(\|\boldsymbol{\pi}^*\|_0 + \|\boldsymbol{\theta}^*\|_0)\log(p \vee n)\right)$, where $\boldsymbol{\gamma}_{h,n} = \boldsymbol{\pi}^* + n^{-1/2}h\boldsymbol{\theta}^*$.*

*Proof of Lemma 11.* Define $\mathbf{V} = \mathbf{Y} - \mathbf{Z}\beta_0$, $\varepsilon_{(h),i} = \mathbf{x}_i^\top \boldsymbol{\mu}^* + n^{-1/2}hu_i + \varepsilon_i$ and $a_{h,\gamma}^2 = n^{-1}\max_{1 \leq j \leq p}\sum_{i=1}^n x_{i,j}^2 \varepsilon_{(h),i}^2$ as well as the following events

$$\mathcal{A}_1 = \{\boldsymbol{\gamma}_{h,n} \text{ lies in the feasible set of the optimization problem (2.1) for } a = a_{h,\gamma}.\}$$

$$\mathcal{A}_2 = \left\{\min_{A \subset \{1,\cdots,p\}, |A| \leq s} \min_{\|v_{A^c}\|_1 \leq \|v_A\|_1} \frac{\|n^{-1/2}\mathbf{X}v\|_2}{\|v_J\|_2} > K\right\}$$

$$\mathcal{A}_3 = \{a_{h,\gamma} \in \mathcal{S}_\gamma\}$$

$$\mathcal{A}_4 = \{\widehat{a}_\gamma \leq 3a_{h,\gamma}\},$$

where $K > 0$ is the constant defined in Lemma 5. By Lemmas 10 and 5,

$$P\left(\mathcal{A}_1 \bigcap \mathcal{A}_2\right) \to 1.$$

Notice that $\|\boldsymbol{\gamma}_{h,n}\|_0 \leq \|\boldsymbol{\pi}^*\|_0 + \|\boldsymbol{\theta}^*\|_0 = o(\sqrt{n}/\log p)$. The rest of the proof proceeds with essentially the same arguments as in the proof of Lemma 7: show $P(\mathcal{A}_3) \to 1$, $P(\mathcal{A}_4) \to 1$ and then the desired result. The only difference is that $(\mathbf{Z}, \boldsymbol{\theta}^*, \widetilde{\boldsymbol{\theta}}(a_{*,\theta}), \widetilde{\boldsymbol{\theta}}(\widehat{a}_\theta), s_\theta)$ is replaced by $(\mathbf{V}, \boldsymbol{\gamma}_{h,n}, \widetilde{\boldsymbol{\gamma}}(a_{h,\gamma}), \widetilde{\boldsymbol{\theta}}(\widehat{a}_\gamma), \|\boldsymbol{\pi}^*\|_0 + \|\boldsymbol{\theta}^*\|_0)$. We omit the details to avoid repetition. $\square$

**Proof of Theorem 8.** Define $\boldsymbol{\gamma}_{h,n} = \boldsymbol{\pi}^* + n^{-1/2}h\boldsymbol{\theta}^*$ and $\boldsymbol{\varepsilon}_{(h)} = (\varepsilon_{(h),1}, \ldots, \varepsilon_{(h),n})^\top$ with $\varepsilon_{(h),i} = \varepsilon_i + n^{-1/2}hu_i + \mathbf{x}_i^\top\boldsymbol{\mu}^*$. Recall that $\mathbf{V} = \mathbf{Y} - \mathbf{Z}\beta_0$. Notice that under $H_{1,h}$ in (3.4),

$$\mathbf{V} = n^{-1/2}h\mathbf{Z} + \mathbf{X}\boldsymbol{\gamma}^* + \boldsymbol{\varepsilon} = \mathbf{X}\boldsymbol{\gamma}_{h,n} + \boldsymbol{\varepsilon}_{(h)}.$$

Let $\widehat{D}^2 = n^{-1}\sum_{i=1}^n \widehat{u}_i^2 \widehat{\varepsilon}_i^2$ and $D^2 = E[(\mathbf{x}_i^\top\boldsymbol{\mu}^* + \varepsilon_i)^2 u_i^2]$, where $\widehat{\varepsilon}_i = v_i - \mathbf{x}_i^\top\widehat{\boldsymbol{\gamma}}$ and $\widehat{u}_i = z_i - \mathbf{x}_i^\top\widehat{\boldsymbol{\theta}}$. Notice that

$$T_n(\beta_0) = \frac{n^{-1/2}(\mathbf{V} - \mathbf{X}\widehat{\boldsymbol{\gamma}})^\top(\mathbf{Z} - \mathbf{X}\widehat{\boldsymbol{\theta}})}{\widehat{D}}. \tag{B.33}$$

The rest of proof proceeds in two steps which establish the behavior of the numerator and the denominator, respectively.

**Step 1: Establish that** $n^{-1/2}(\mathbf{V} - \mathbf{X}\widehat{\boldsymbol{\gamma}})^\top(\mathbf{Z} - \mathbf{X}\widehat{\boldsymbol{\theta}})/D \to^d N(h\kappa, 1)$.

Observe that

$$n^{-1/2}(\mathbf{V} - \mathbf{X}\widehat{\boldsymbol{\gamma}})^\top(\mathbf{Z} - \mathbf{X}\widehat{\boldsymbol{\theta}}) = \underbrace{n^{-1/2}\boldsymbol{\varepsilon}_{(h)}^\top \mathbf{u}}_{A_1} + \underbrace{n^{-1/2}\boldsymbol{\varepsilon}_{(h)}^\top \mathbf{X}(\boldsymbol{\theta}^* - \widehat{\boldsymbol{\theta}})}_{A_2} + \underbrace{n^{-1/2}(\boldsymbol{\gamma}_{h,n} - \widehat{\boldsymbol{\gamma}})^\top \mathbf{X}^\top(\mathbf{Z} - \mathbf{X}\widehat{\boldsymbol{\theta}})}_{A_3}. \tag{B.34}$$

Notice that

$$|A_3| \leq \|n^{1/2}(\boldsymbol{\gamma}_{h,n} - \widehat{\boldsymbol{\gamma}})\|_1 \|n^{-1}\mathbf{X}^\top(\mathbf{Z} - \mathbf{X}\widehat{\boldsymbol{\theta}})\|_\infty$$
$$\stackrel{(i)}{=} O_P\left(n^{-1/2}(\|\boldsymbol{\pi}^*\|_0 + \|\boldsymbol{\theta}^*\|_0)\log(p \vee n)\right) \|n^{-1}\mathbf{X}^\top(\mathbf{Z} - \mathbf{X}\widehat{\boldsymbol{\theta}})\|_\infty$$
$$\stackrel{(ii)}{\leq} O_P\left(n^{-1/2}(\|\boldsymbol{\pi}^*\|_0 + \|\boldsymbol{\theta}^*\|_0)\log(p \vee n)\right) O_P(\log(p \vee n)) = o_P(1), \tag{B.35}$$

where $(i)$ follows by Lemma 11 and $(ii)$ follows by Lemma 7.

For $A_2$, we observe that

$$|A_2| \leq \|n^{-1/2}\mathbf{X}^\top\boldsymbol{\varepsilon}_{(h)}\|_\infty \|(\boldsymbol{\theta}^* - \widehat{\boldsymbol{\theta}})\|_1$$
$$\stackrel{(i)}{=} O_P(\log(p \vee n))\|(\boldsymbol{\theta}^* - \widehat{\boldsymbol{\theta}})\|_1$$
$$\stackrel{(ii)}{=} O_P(\log(p \vee n)) O_P(n^{-1/2}s_\theta \log(p \vee n)) = o_P(1), \tag{B.36}$$

where $(i)$ follows by the second result in Lemma 10 and $(ii)$ follows by Lemma 7.

For $A_1$, we notice that

$$A_1 = n^{-1/2}\sum_{i=1}^n \varepsilon_{(h),i} u_i = hn^{-1}\sum_{i=1}^n u_i^2 + n^{-1/2}\sum_{i=1}^n u_i(\mathbf{x}_i^\top\boldsymbol{\mu}^* + \varepsilon_i)$$
$$\stackrel{(i)}{=} hE(u_1^2) + o_P(1) + n^{-1/2}\sum_{i=1}^n u_i(\mathbf{x}_i^\top\boldsymbol{\mu}^* + \varepsilon_i), \tag{B.37}$$

where $(i)$ follows by the law of large numbers. Now we combine (B.35), (B.36) and (B.37) with (B.34) to obtain

$$\frac{n^{-1/2}(\mathbf{V} - \mathbf{X}\widehat{\boldsymbol{\gamma}})^\top(\mathbf{Z} - \mathbf{X}\widehat{\boldsymbol{\theta}})}{D} = \frac{hE(u_1^2) + o_P(1) + n^{-1/2}\sum_{i=1}^n u_i(\mathbf{x}_i^\top\boldsymbol{\mu}^* + \varepsilon_i)}{D}$$
$$\stackrel{(i)}{=} h\kappa + \frac{n^{-1/2}\sum_{i=1}^n u_i(\mathbf{x}_i^\top\boldsymbol{\mu}^* + \varepsilon_i)}{D} + o_P(1)$$
$$\stackrel{(ii)}{\to}^d N(h\kappa, 1), \tag{B.38}$$

where $(i)$ follows by the fact that $D$ is bounded away from zero and the assumption that $E(u_1^2)/D \to \kappa$ and $(ii)$ follows by the central limit theorem.

**Step 2: Establish that $\widehat{D}/D = 1 + o_P(1)$.**

Since $\widehat{u}_i - u_i = \mathbf{x}_i^\top(\boldsymbol{\theta}^* - \widehat{\boldsymbol{\theta}})$, we have that

$$\begin{aligned}
\max_{1 \leq i \leq n} |\widehat{u}_i^2 - u_i^2| &= \max_{1 \leq i \leq n} |\widehat{u}_i + u_i| \cdot |\mathbf{x}_i^\top(\widehat{\boldsymbol{\theta}} - \boldsymbol{\theta}^*)| \\
&\leq \max_{1 \leq i \leq n} \left(|2u_i| + |\mathbf{x}_i^\top(\widehat{\boldsymbol{\theta}} - \boldsymbol{\theta}^*)|\right) |\mathbf{x}_i^\top(\widehat{\boldsymbol{\theta}} - \boldsymbol{\theta}^*)| \\
&\leq (2\|\mathbf{u}\|_\infty + \|\mathbf{X}\|_\infty \|\widehat{\boldsymbol{\theta}} - \boldsymbol{\theta}^*\|_1) \|\mathbf{X}\|_\infty \|\widehat{\boldsymbol{\theta}} - \boldsymbol{\theta}^*\|_1 \\
&\stackrel{(i)}{\leq} O_P\left(n^{-1} s_\theta^2 \log^3(p \vee n)\right),
\end{aligned} \tag{B.39}$$

where $(i)$ follows by Lemma 7 and the sub-Gaussian property of $u_i$ and $x_i$ (which, by the union bound, implies $\|\mathbf{u}\|_\infty = O_P(\sqrt{\log n})$ and $\|\mathbf{X}\|_\infty = O_P(\sqrt{\log(pn)})$). Similarly, using Lemma 11, we can obtain

$$\max_{1 \leq i \leq n} |\widehat{\varepsilon}_{(h),i}^2 - \varepsilon_{(h),i}^2| = O_P\left(n^{-1}(\|\pi^*\|_0 + \|\mu^*\|_0)^2 \log^3(p \vee n)\right). \tag{B.40}$$

Now we observe

$$\begin{aligned}
&\left|n^{-1} \sum_{i=1}^n (\widehat{\varepsilon}_{(h),i}^2 \widehat{u}_i^2 - \varepsilon_{(h),i}^2 u_i^2)\right| \\
&\leq \left(\max_{1 \leq i \leq n} \widehat{u}_i^2\right) \left|n^{-1} \sum_{i=1}^n (\widehat{\varepsilon}_{(h),i}^2 - \varepsilon_{(h),i}^2)\right| + \left(\max_{1 \leq i \leq n} \varepsilon_{(h),i}^2\right) \left|n^{-1} \sum_{i=1}^n (\widehat{u}_i^2 - u_i^2)\right| \\
&\leq \left(\max_{1 \leq i \leq n} \widehat{u}_i^2\right) \left(\max_{1 \leq i \leq n} \left|\widehat{\varepsilon}_{(h),i}^2 - \varepsilon_{(h),i}^2\right|\right) + \left(\max_{1 \leq i \leq n} \varepsilon_{(h),i}^2\right) \left(\max_{1 \leq i \leq n} |\widehat{u}_i^2 - u_i^2|\right) \\
&\leq \left(\max_{1 \leq i \leq n} u_i^2 + \max_{1 \leq i \leq n} |\widehat{u}_i^2 - u_i^2|\right) \left(\max_{1 \leq i \leq n} \left|\widehat{\varepsilon}_{(h),i}^2 - \varepsilon_{(h),i}^2\right|\right) \\
&\quad + \left(\max_{1 \leq i \leq n} \varepsilon_{(h),i}^2\right) \left(\max_{1 \leq i \leq n} |\widehat{u}_i^2 - u_i^2|\right) \\
&\stackrel{(i)}{=} [O_P(\log n) + O_P(n^{-1} s_\theta^2 \log^3(p \vee n))] O_P(n^{-1}(\|\pi^*\|_0 + \|\mu^*\|_0)^2 \log^3(p \vee n)) \\
&\quad + O_P(\log n) O_P(n^{-1} s_\theta^2 \log^3(p \vee n)) \\
&= o_P(1),
\end{aligned} \tag{B.41}$$

$$\tag{B.42}$$

where $(i)$ follows by (B.39) and (B.40) together with $\max_i u_i^2 = O_P(\log n)$ and $\max_i \varepsilon_{(h),i}^2 = O_P(\log n)$ (due to the sub-Gaussian norms of $u_i$ and $\varepsilon_{(h),i}$ and the union bound). The law of large numbers implies that

$$\begin{aligned}
n^{-1} \sum_{i=1}^n \varepsilon_{(h),i}^2 u_i^2 &= o_P(1) + E\varepsilon_{(h),1}^2 u_1^2 \\
&= o_P(1) + Eu_1^2 (\mathbf{x}_i^\top \boldsymbol{\mu}^* + \varepsilon_i)^2 + n^{-1} h^2 Eu_1^2 = o_P(1) + D.
\end{aligned}$$

Therefore, (B.42) and the above display imply that

$$\widehat{D}^2 - D^2 = n^{-1} \sum_{i=1}^n (\widehat{\varepsilon}_{(h),i}^2 \widehat{u}_i^2 - \varepsilon_{(h),i}^2 u_i^2) + n^{-1} \sum_{i=1}^n \varepsilon_{(h),i}^2 u_i^2 - D = o_P(1).$$

Since $D$ is bounded away from zero, we have

$$\frac{\widehat{D}}{D} = \frac{\sqrt{D^2 + o_P(1)}}{D} = 1 + o_P(1). \tag{B.43}$$

Now we combine (B.38) and (B.43) with (B.33) to obtain that

$$T_n(\beta_0) \to^d N(h\kappa, 1).$$

Therefore,

$$P\left(|T_n(\beta_0)| > \Phi^{-1}(1 - \alpha/2)\right) \to P\left(|h\kappa + \zeta| > \Phi^{-1}(1 - \alpha/2)\right),$$

where $\zeta \sim N(0, 1)$. By elementary computations, we have

$$P\left(|h\kappa + \zeta| > \Phi^{-1}(1 - \alpha/2)\right) = \Psi(h, \kappa, \alpha).$$

The proof is complete.

□